\documentclass[superscriptaddress,prx,twocolumn]{revtex4-2}
\usepackage{graphicx}
\usepackage{amsmath,amssymb,physics,dsfont}
\usepackage{mathrsfs}
\usepackage{color}
\usepackage{hyperref}
\usepackage{microtype}
\usepackage{braket}
\usepackage{stmaryrd}
\usepackage{academicons}
\usepackage{defs-private}
\usepackage{qcircuit}
\usepackage{booktabs}
\usepackage[dvipsnames]{xcolor}

\begin{document}

\title{Quantum-Classical Embedding via Ghost Gutzwiller Approximation for Enhanced Simulations of Correlated Electron Systems}

\author{I-Chi Chen}
\affiliation{Ames National Laboratory, U.S. Department of Energy, Ames, Iowa 50011, USA}
\affiliation{Department of Physics and Astronomy, Iowa State University, Ames, Iowa 50011, USA}

\author{Aleksei Khindanov}
\affiliation{Ames National Laboratory, U.S. Department of Energy, Ames, Iowa 50011, USA}

\author{Carlos Salazar}
\affiliation{School of Electrical Engineering and Computer Science, Washington State University, Pullman, Washington 99163, USA}

\author{Humberto Munoz Barona}
\affiliation{Department of Mathematics, Southern University and A$\&$M College, Baton Rouge, LA 70807, USA}

\author{Feng Zhang}
\affiliation{Ames National Laboratory, U.S. Department of Energy, Ames, Iowa 50011, USA}
\affiliation{Department of Physics and Astronomy, Iowa State University, Ames, Iowa 50011, USA}

\author{Cai-Zhuang Wang}
\affiliation{Ames National Laboratory, U.S. Department of Energy, Ames, Iowa 50011, USA}
\affiliation{Department of Physics and Astronomy, Iowa State University, Ames, Iowa 50011, USA}

\author{Thomas Iadecola}
\affiliation{Ames National Laboratory, U.S. Department of Energy, Ames, Iowa 50011, USA}
\affiliation{Department of Physics and Astronomy, Iowa State University, Ames, Iowa 50011, USA}

\author{Nicola Lanat\`a}
\affiliation{School of Physics and Astronomy, Rochester Institute of Technology, Rochester, New York 14623, USA}
\affiliation{Center for Computational Quantum Physics, Flatiron Institute, New York, New York 10010, USA}

\author{Yong-Xin Yao}
\email{ykent@iastate.edu}
\affiliation{Ames National Laboratory, U.S. Department of Energy, Ames, Iowa 50011, USA}
\affiliation{Department of Physics and Astronomy, Iowa State University, Ames, Iowa 50011, USA}

\begin{abstract}
Simulating correlated materials on present-day quantum hardware remains challenging due to limited quantum resources. Quantum embedding methods offer a promising route by reducing computational complexity through the mapping of bulk systems onto effective impurity models, allowing more feasible simulations on pre- and early-fault-tolerant quantum devices. This work develops a quantum-classical embedding framework based on the ghost Gutzwiller approximation to enable quantum-enhanced simulations of ground-state properties and spectral functions of correlated electron systems. Circuit complexity is analyzed using an adaptive variational quantum algorithm on a statevector simulator, applied to the infinite-dimensional Hubbard model with increasing ghost mode numbers from 3 to 5, resulting in circuit depths growing from 16 to 104. Noise effects are examined using a realistic error model, revealing significant impact on the spectral weight of the Hubbard bands. To mitigate these effects, the Iceberg quantum error detection code is employed, achieving up to 40\% error reduction in simulations. Finally, the accuracy of the density matrix estimation is benchmarked on IBM and Quantinuum quantum hardware, featuring distinct qubit-connectivity and employing multiple levels of error mitigation techniques. 
\end{abstract}

\maketitle

\section{Introduction}
Recent advancements in quantum computing have opened new avenues for simulating complex materials, yet significant challenges remain in terms of resource requirements. 
Fault-tolerant quantum computing necessitates the implementation of error correction protocols, which substantially increase the number of physical qubits and quantum gates needed for accurate computations. 
Current estimates for quantum resource requirements in bulk materials simulations, particularly those leveraging the leading quantum phase estimation algorithm with qubitization, indicate a need for more than \(10^6\) physical qubits and over \(10^{10}\) Toffoli gates~\cite{Ivanov2023QuantumCP, ivanov2024quantum, berry2024quantum, georges2024quantum}. 
For near-term quantum simulations of bulk materials, rigorous resource estimation is hampered by the heuristic nature of the associated quantum algorithms, such as the variational quantum eigensolver~\cite{vqe_theory}. 
A recent partial estimate, accounting for only a single layer of a variational Hamiltonian ansatz with circuit complexity comparable to one first-order Trotter step, suggests that simulations may require hundreds to thousands of physical qubits, thousands to millions of two-qubit gates, or circuit depths ranging from thousands to tens of thousands~\cite{clinton2024towards}. 
These estimates are based on reduced models with a limited number of low-energy bands. 
While significantly lower than the requirements for fault-tolerant approaches, these resource demands still far exceed the capabilities of current quantum hardware.

Quantum embedding methods offer a strategy to reduce quantum resource requirements by mapping bulk system simulations onto smaller fragments subject to self-consistency conditions. 
Among these, Dynamical Mean-Field Theory (DMFT) is a prominent approach derived variationally from the Baym-Kadanoff functional~\cite{dmft_georges96, dmft_held07}. 
DMFT maps a lattice correlated electron system onto an effective impurity model with a self-consistency condition, and becomes exact in the limit of infinite spatial dimensions. 
Efforts have been devoted to integrating DMFT with quantum computing by employing quantum hardware as a scalable impurity solver~\cite{hybrd_dmft}. 
The required quantum resources are nevertheless still substantial. 
A precise discretization of the continuous bath (consisting of degrees of freedom outside the impurity) often requires a large number of bath sites, significantly increasing the complexity of the impurity model.
Additionally, computing the one-particle Green’s function, a key quantity in DMFT, involves multiple computationally intensive steps, including ground state preparation, real-time evolution for state propagation, and the measurement of state overlaps~\cite{hybrd_dmft}.
This leads to an estimation that circuits with \(\sim 10^8\) gates are needed for impurity models with 10 spatial orbitals plus 60-100 bath sites.
Nevertheless, several demonstrations of DMFT calculations utilizing quantum computing as an impurity solver have been reported~\cite{hybrd_2sitedmft2, Backes2023DMFT, selisko2024DMFT}. These studies employ approximate bath discretizations with a limited number of bath sites and leverage near-term quantum algorithms to compute the Green’s function, making the approach more feasible for current quantum hardware.

The Gutzwiller variational embedding approach based on the Gutzwiller approximation (GA) offers an efficient alternative to DMFT by requiring only the ground state solution of an impurity model with minimal bath sites, significantly reducing computational complexity. 
It has been shown to yield ground-state properties comparable to those obtained through DMFT in simulations of real materials, including transition metals, rare-earth elements, and actinide compounds~\cite{ga_ce, deng2008lda+, ga_pu, ga_uo2, ga_tmo}. 
The Gutzwiller Quantum-Classical Embedding (GQCE) framework has been developed to integrate quantum computing as the impurity solver for ground state calculations~\cite{gqce}. 
Full self-consistent GQCE simulations on quantum hardware have been benchmarked in studies of the phase diagram of the periodic Anderson model on a Bethe lattice~\cite{gqce}. 
Extensions to correlated multi-band systems have been explored, focusing on ground state preparation of multi-orbital impurity models using adaptive variational quantum algorithms~\cite{mukherjee2023comparative}. 
Density Matrix Embedding Theory (DMET) is another widely used embedding approach for ground state calculations. 
Recent studies have explored the connection between DMET and GA~\cite{ga_pu, ga_dmet}. 
DMET has also been extended to quantum computing in a hybrid approach, where a quantum computer is employed as an impurity solver to calculate the ground state properties of electron systems~\cite{Mineh2022DMET, kawashima2021DMET, rubin2016DMET}.

In this work, we develop an extended version of the GQCE approach, abbreviated as GQCE(g), by incorporating the Ghost Gutzwiller Approximation (gGA)~\cite{gga_lanata2017, Frank2021QuantumED, Lee2024ChargeSC}. 
It introduces ghost orbitals into the noninteracting quasi-particle wavefunction, thereby enlarging the variational space. 
This leads to a substantial improvement not only in ground state properties but also in spectral function calculations. 
By increasing the number of ghost orbitals, gGA calculations converge to the accuracy of DMFT. 
The approach preserves computational efficiency relative to DMFT by leveraging the ground state solution of an impurity model with fewer bath sites.
Nevertheless, compared to GQCE, the impurity model in GQCE(g) becomes larger due to the inclusion of ghost orbitals. 
To assess the feasibility of ground state preparation, we employ the adaptive variational quantum imaginary time evolution (AVQITE) algorithm~\cite{AVQITE} on a statevector simulator, benchmarking impurity models with an increasing number of bath sites in GQCE(g) calculations for the single-band Hubbard model on a Bethe lattice.
Furthermore, using a realistic noise model, we analyze the impact of quantum noise on key physical quantities, including the spectral function and self-energy. 
To mitigate errors, we demonstrate that quantum error detection (QED) codes can enhance the accuracy of observable expectation values for the impurity model. 
Finally, we assess the feasibility of measuring expectation values on IBM and Quantinuum quantum processing units (QPUs), focusing on the effects of qubit connectivity, and circuit complexity and error mitigation techniques.

\begin{figure}[ht]
	\centering
	\includegraphics[width=\linewidth]{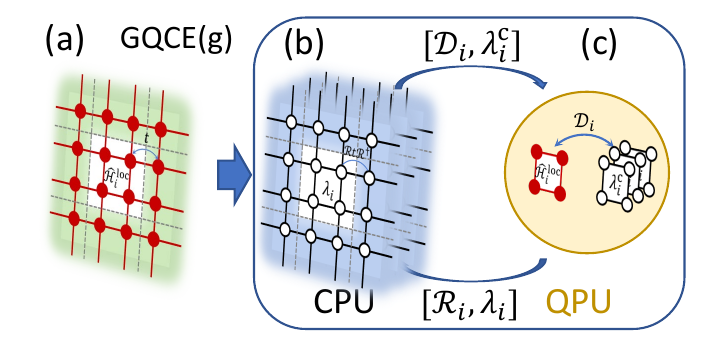}
	\caption{
	\textbf{Schematic illustration of the GQCE(g) framework.} (a) Illustration of an interacting fermionic lattice model, exemplified by a unit cell containing 4 orbital sites. The unit cell is treated as a single $i=1$ group, with the local part of the system Hamiltonian denoted by $\h_i^\text{loc}$ and the inter-cell hopping by $t$. (b, c) Within the GQCE(g) framework,  the ground state solution of the interacting lattice model is mapped to a self-consistent solution of a noninteracting quasiparticle lattice model (b) to a finite-size, interacting embedding impurity model (c). The quasiparticle lattice model repeats a unit cell of $B(=3)\times4$ auxiliary orbital sites, with intra-cell one-body Hamiltonian term given by $\lambda_i$, and inter-cell hoppings by $\R_i t \R_i^\dag$. The embedding model is composed of $\h_i^\text{loc}$ defined on the $4$ physical sites and the $B(=3)\times4$ bath sites described by $\lambda_i^\mathrm{c}$, with the mutual coupling strength $\D_i$.  Within GQCE(g), the ground state of the embedding Hamiltonian is solved on QPUs using (adaptive variational) quantum algorithms such as AVQITE. The quasiparticle Hamiltonian can be efficiently simulated on classical processing units (CPUs).
	}
	\label{fig: gqceg}
\end{figure}

\section{General exposition of the GQCE(g) framework}
Consider the following generic multi-orbital interacting fermionic model on a periodic lattice:
\be
\h = \sum_{I,J=1}^{\mathcal{N}}\sum_{i,j=0}^{\mathcal{C}}\sum_{\alpha=1}^{\nu_i}\sum_{\beta=1}^{\nu_j} t_{IiJj}^{\alpha \beta}\, \cc_{Ii\alpha}\ca_{Jj \beta} +\sum_{I=1}^{\mathcal{N}}\sum_{i=1}^{\mathcal{C}}\h_{Ii}^\text{loc}, \label{eq: model}
\ee
where $I,J$ run through $\mathcal{N}$ unit cells. For convenience and generality, we partition the orbital-sites of a unit cell into $\mathcal{C}+1$ groups labeled by $i,j$. The $0$th group contains all ``uncorrelated'' orbitals, which typically correspond to delocalized $s$- or $p$-orbitals. The remaining $\mathcal{C}$ groups contain ``correlated'' atomic orbitals with substantial screened Coulomb interactions, which typically correspond to local $d$- or $f$-orbitals for transition-metal or rare-earth/actinide compounds. Specifically, an $i\geq 1$ group can contain $5$ $d$-orbitals at one transition-metal site, or all $d$-orbitals at multi-atomic sites in the unit cell. The former setup is often used for single-site quantum embedding, while the latter is used for cluster embedding. The $\nu_i$ spin-orbitals in the $i$th group are indexed by $\alpha$ or $\beta$. The hopping matrix elements are given by $t_{IiJj}^{\alpha \beta}$, which are associated with the annihilation ($\ca_{Jj \beta}$) and creation ($\cc_{Ii\alpha}$) operators. We set $t_{IiIi}^{\alpha \beta} \equiv 0$ for all the $\mathcal{C}$ ``correlated'' groups, and isolate all the Hamiltonian components defined on the spin-orbitals of the group $i\geq 1$ to a local part $\h_{Ii}^\text{loc}$, which can include one-body terms (e.g., crystal field and spin-orbit coupling) and two-body Coulomb interactions.

To simulate the ground state of the model~\eqref{eq: model}, we adopt the generalized Gutzwiller wavefunction:
\be
\ket{\Psi_\text{G}}=\hat{\mathcal{P}}_\text{G}\ket{\Psi_0}\,, \label{eq: gwf}
\ee
where $\ket{\Psi_0}$ is a single-particle reference state. The operator $\hat{\mathcal{P}}_\text{G}=\Pi_{I=1}^{\mathcal{N}}\Pi_{i=1}^{\mathcal{C}}\hat{\mathcal{P}}_{Ii}$ involves local operators defined as:
\be
\hat{\mathcal{P}}_{Ii} = \sum_{\Gamma = 0}^{2^{\nu_i} - 1} \sum_{n = 0}^{2^{B \nu_i} - 1}\left[\Lambda_i\right]_{\Gamma n} \ket{\Gamma, Ii}\bra{n, Ii}\,,
\ee
where $\ket{\Gamma, Ii} = \Pi_{\alpha=1}^{\nu_i} (\cc_{Ii\alpha})^{q_\alpha(\Gamma)}\ket{0}$ labels the local Fock states defined on the $\nu_i$ fermionic $c$-modes, while $\ket{n, Ii} = \Pi_{a=1}^{B\nu_i} (\fc_{Ii\alpha})^{q_\alpha(n)}\ket{0}$ represents Fock states generated on the $B\nu_i$ fermionic $f$-modes. Here $q_\alpha(\Gamma)$ denotes the occupation of the $\alpha$th spin-orbital in the Fock state $\ket{\Gamma, Ii}$, which corresponds to the $\alpha$th digit in the binary representation of the integer $\Gamma$. The number of ghost modes in the wavefunction~\eqref{eq: gwf} is set by the bath orbital parameter $B$, where the conventional Gutzwiller wavefunction is restored with $B = 1$, and $(B-1)\nu_i$ ghost modes are introduced with $B > 1$. The additional variational degrees of freedom introduced in $\ket{\Psi_\text{G}}$ are encoded in the $2^{\nu_i} \times 2^{B\nu_i}$ complex matrix $[\Lambda_i]_{\Gamma n}$, where the nonzero elements are determined by the symmetry of the represented phase. For the normal phase with $U(1)$ symmetry (i.e., particle number conservation), the following selection rule holds:
\be
\sum_{a=1}^{B\nu_i}q_a(n) - \sum_{\alpha=1}^{\nu_i}q_{\alpha}(\Gamma) = m_i \quad \forall \Gamma, n \,|\, \left[\Lambda_i\right]_{\Gamma n} \neq 0 \,.
\ee
As a result, the total number of physical particles in $\ket{\Psi_\text{G}}$ is less than the number of auxiliary particles in $\ket{\Psi_0}$ by an amount $\mathcal{N}\sum_{i=1}^{\mathcal{C}}m_i$. Note that, to reduce $\ket{\Psi_{\rm G}}$ to the standard Gutzwiller wavefunction, one needs to set $B=1$, which yields $m_i = 0$. For $B>1$, the integer $m_i$ can be optimized in principle. Empirically, we can assume $B$ is odd and $m_i = (B-1)\nu_i/2$, which sets the $(B-1)\nu_i$ ghost modes at half-filling.
From this point onward, we omit the unit cell index $I$ for notations that are translationally invariant, such as the $[\Lambda_i]_{\Gamma n}$ matrix.

The direct optimization of the total energy $\mathcal{E}(\Psi_0, \hat{\mathcal{P}}_\text{G}) = \avihi{\Psi_\text{G}}{\h}$ can be challenging due to the large number of variational degrees of freedom in $\Psi_0$ and $\hat{\mathcal{P}}$. To reduce the computational complexity, we resort to the ghost Gutzwiller approximation and get the following closed form for the total energy:
\be
\mathcal{E} \approx \avihi{\Psi_0}{\hat{\T}_\text{G}} +\mathcal{N}\sum_{i=1}^{\mathcal{C}}\avihi{\Phi_{i}}{\h_{i}^\text{loc}}\,, \label{eq: e}
\ee
with 
\be
\hat{\T}_\text{G} = \sum_{I,J=1}^{\mathcal{N}}\sum_{i,j=0}^{\mathcal{C}}\sum_{a=1}^{B\nu_i}\sum_{b=1}^{B\nu_j} \sum_{\alpha=1}^{\nu_i}\sum_{\beta=1}^{\nu_j} \R_{ia \alpha} t_{IiJj}^{\alpha, \beta}\R_{j\beta b}^\dag\,\fc_{Iia}\fa_{Jjb} \,.
\ee
Here $a, b$ denote the fermionic $f$-modes, which include both spatial and spin components unless spin is explicitly labeled. The above expression assumes the Gutzwiller constraints:
\be
    \begin{aligned}
        \ovii{\Phi_i} &= 1\,, \\
        \avihi{\Phi_i}{\ba_{i b}\bc_{i a}} &= \avihi{\Psi_0}{\fc_{i a}\fa_{i b}} \equiv [\Delta_i]_{ab}\,.
    \end{aligned}
 \label{eq: gc}
\ee
The matrix $\R_i$ is an identity for $i=0$, and a $B\nu_i \times \nu_i$ matrix defined through the following equation for $i\geq 1$:
\be
\avihi{\Phi_i}{\cc_{i\alpha}\ba_{i a}} = \left[\R_i^\text{T}\sqrt{\Delta_i(1-\Delta_i)} \right]_{\alpha a}\, , \label{eq: R}
\ee
where $\ket{\Phi_i}$ is a many-body state in the local Hilbert space defined by the $\nu_i$ fermionic $f$-modes (physical sites) and $B\nu_i$ fermionic $b$-modes (bath sites).

The minimization of the total energy~\eqref{eq: e} subject to the Gutzwiller constraints~\eqref{eq: gc} leads to the following coupled eigenvalue equations:
\bea
\h^\text{qp}\left[\R, \lambda\right]\ket{\Psi_0} &= E_0\ket{\Psi_0} \,, \label{eq: eqp} \\ 
\h^\text{eb}_i\left[\D_i, \lambda_i^\mathrm{c}\right]\ket{\Phi_i} &= E_i\ket{\Phi_i} \,,  \label{eq: eeb}
\eea
for the noninteracting quasi-particle Hamiltonian:
\be
\h^\text{qp}\left[\R, \lambda\right] = \hat{\T}_\text{G} + \sum_{I=1}^{\mathcal{N}}\sum_{i=1}^{\mathcal{C}} \sum_{a,b=1}^{B\nu_i}\left[\lambda_i\right]_{ab}\fc_{Iia}\fa_{Iib} \,, \label{eq: hqp}
\ee
and the many-body embedding Hamiltonian:
\bea
\h^\text{eb}_i\left[\D_i, \lambda_i^\mathrm{c} \right] &=& \h^\text{loc}_i\left[\ca_{i\alpha}, \cc_{i\alpha}\right] + \sum_{a,b=1}^{B\nu_i}\left[\lambda_i^\mathrm{c} \right]_{ab}\ba_{ib}\bc_{ia} \nonumber \label{eq: heb} \\
&&+ \sum_{a=1}^{B\nu_i}\sum_{\alpha=1}^{\nu_i}\left(\left[\D_i\right]_{a\alpha}\cc_{i\alpha}\ba_{ia}  + H.c.\right) 
\eea
with $i=1, \dots, \mathcal{C}$. The $B\nu_i \times \nu_i$ matrix $\D_i$ is defined though the equation:
\be
\begin{aligned}
\sum_{J=1}^{\mathcal{N}}\sum_{j=0}^{\mathcal{C}}\sum_{b=1}^{B\nu_j} \sum_{\beta=1}^{\nu_j} t_{1iJj}^{\alpha, \beta}\R_{j\beta b}^\dag\, &\avihi{\Psi_0}{\fc_{1ia}\fa_{Jjb}} \\
=& \left[\sqrt{\Delta_i(1-\Delta_i)} \D_i \right]_{a \alpha}.    
\end{aligned}
\ee
Note that $I=1$ for $t_{1iJj}^{\alpha, \beta}$ and $\fc_{1ia}$ refers to the central unit cell. The matrix $\lambda^\mathrm{c}_i$ is related to $\lambda_i$ through the following equation:
\be
\frac{\partial \left[\Tr{\R_i^\text{T}\sqrt{\Delta_i(1-\Delta_i)}\D_i} + c.c.\right]}{\partial \left[\Delta_i\right]_{ab}} 
+ \left[\lambda_i\right]_{ab}+\left[\lambda_i^\mathrm{c}\right]_{ab} = 0.    
\ee
In practice, ground states of the coupled Hamiltonian systems~(\ref{eq: eqp},~\ref{eq: eeb}) are iteratively solved until self-consistency (e.g., in $\R$ and $\lambda$) is achieved.

Since the eigenvalue problem of the noninteracting quasiparticle Hamiltonian~\eqref{eq: eqp} can be efficiently solved using classical algorithms, the computational bottleneck lies in the ground state solution of the many-body embedding Hamiltonian~\eqref{eq: eeb}, where the classical computational resources can scale exponentially with the number of single-particle modes. Within the GQCE(g) framework, as schematically summarized in Fig.~\ref{fig: gqceg}, for the embedding Hamiltonian we resort to quantum algorithms executed on QPUs, where the number of qubits needed to represent the quantum state grows only linearly with system size. Specifically for pre-fault-tolerant applications, we adopt the adaptive variational quantum imaginary time evolution algorithm (AVQITE) for efficient ground state preparation~\cite{AVQITE, AVQDS}.

\section{Application to the Hubbard model on the Bethe lattice}
To demonstrate the GQCE(g) approach, we choose a specific single-band Hubbard model on the Bethe lattice represented as:
\be
\h = \sum_{\braket{I,J}}\sum_{\sigma}t \cc_{I\sigma}\ca_{J\sigma} + \sum_{I} \frac{U}{2} (\hat{n}_I - 1)^2\,. \label{eq: sbhm}
\ee
Here $U$ is the onsite Hubbard interaction strength and $\hat{n}_I = \sum_{\sigma=\uparrow, \downarrow} \cc_{I\sigma}\ca_{I\sigma}$ is the onsite electron occupation number operator. Compared with the generic fermionic model~\eqref{eq: model}, this specific Hamiltonian~\eqref{eq: sbhm} is free of ``uncorrelated'' orbitals and has only one group of ``correlated'' orbitals in each unit cell. Therefore, the summation over groups $i, j$ in Eq.~\eqref{eq: model} is omitted and the summation over spin-orbitals $\alpha, \beta$ is reduced to that over the spin index $\sigma$. The notation $\braket{I,J}$ implies a summation over nearest neighbor unit cell sites, where the coordination number of each site is infinite for the Bethe lattice. In practice, the first term in the Hamiltonian~\eqref{eq: sbhm} with nearest-neighbor hopping amplitude $t$ as the noninteracting component is represented by a density of states in the semicircular form $D_0(\epsilon) = \frac{2}{\pi W}\sqrt{1-(\epsilon/W)^2}$, where the half-bandwidth $W$ is hereafter set to be 1 as the energy unit. The form of the second term in the Hamiltonian~\eqref{eq: sbhm} ensures that the ground state is at half-filling, i.e., the expectation value $\braket{\hat{n}_I} = 1$. 

When applying GQCE(g) to the model~\eqref{eq: model} the embedding Hamiltonian~\eqref{eq: heb} reduces to the following form:
\be
\begin{aligned}
\h^\text{eb}&\left[\D, \lambda^\mathrm{c} \right] = U\cc_{\uparrow}\ca_{\uparrow}\cc_{\downarrow}\ca_{\downarrow} -\frac{U}{2}\sum_{\sigma=\uparrow, \downarrow} \cc_{\sigma}\ca_{\sigma} \\
&+ \sum_{a=1}^{B}\sum_{\sigma=\uparrow, \downarrow} \left[\lambda^\mathrm{c} \right]_{a\sigma}\ba_{a\sigma}\bc_{a\sigma} + (\left[\D\right]_{a\sigma}\cc_{\sigma}\ba_{a\sigma} + H.c.)\,. \label{eq: heb1}    
\end{aligned} 
\ee
The above expression adopts a gauge where the $\lambda_\mathrm{c}$ is diagonal~\cite{Frank2024activeLA}. For ground state preparation using quantum algorithms, we apply the standard Jordan-Wigner encoding to transform the Hamiltonian~\eqref{eq: heb1} from a fermionic to a qubit representation, where the number of qubits used is the same as the number of spin-orbitals. In the GQCE(g) calculation for the paramagnetic phase of the model~\eqref{eq: sbhm}, the embedding Hamiltonian~\eqref{eq: heb1} is solved for the lowest-energy spin-singlet state $\ket{\Phi}$ at half-filling, i.e., $\sum_{\mu\sigma}\rho_{\mu\mu, \sigma}=B+1$, where the one-particle density matrix $\rho_{\mu\nu, \sigma}$ is defined as:
\be 
\rho_{\mu\nu, \sigma} = \avihi{\Phi}{\aac_{\mu\sigma}\aaa_{\nu\sigma}}, \text{ with } \aaa_{\mu\sigma} \in \setof{\ca_{\sigma}} \cup \setof{ \ba_{a\sigma} }_{a=1}^{B}\,. \label{eq: 1pdm}
\ee
To guarantee that $\ket{\Phi}$ can be obtained from the ground state solution with the required particle number ($B+1$) and spin symmetry ($S=0$), a penalty term involving the total spin operator $\hat{S}$, $g\hat{S}^2 + g(\sum_{\mu\sigma} \rho_{\mu\mu, \sigma} - B-1)^2$, is added to the embedding Hamiltonian~\eqref{eq: heb1} with a fixed prefactor $g=10$.

Due to the spin $\mathbb{Z}_2$ symmetry of this model, in the following calculations we focus on the one-particle density matrix $\rho_{\mu\nu, \sigma}$ from the embedding Hamiltonian ground state solution~\eqref{eq: eeb}, the renormalization vector $\R_{a\uparrow}$~\eqref{eq: R}, the double occupancy $\avihi{\Phi}{\cc_{\uparrow}\ca_{\uparrow}\cc_{\downarrow}\ca_{\downarrow}}$, and the spectral function $A_{\uparrow}(\omega) = -\frac{1}{\pi} \Im{G_{\uparrow}(\omega)}$ which characterizes single-particle excitations. The Green's function defined for the effective noninteracting quasi-particle Hamiltonian~\eqref{eq: hqp} can be evaluated as $G_{\uparrow}(\omega) = \sum_{\epsilon=-1}^{1} D_0(\epsilon) G_{\epsilon\uparrow}(\omega)$ with:
\be
G_{\epsilon\uparrow}(\omega) = \sum_{a, b =1}^{B} \R^{*}_{a\uparrow} \avihi{0}{ \fa_{a\uparrow} \frac{1}{\omega-\h^\text{qp}_{\epsilon}[\R, \lambda] +i 0^+} \fc_{b\uparrow}}\R_{b \uparrow} \,.
\ee
Here $\ket{0}$ is the particle vacuum state and the chemical potential is fixed at $0$. The quasi-particle Hamiltonian~\eqref{eq: hqp} has the following specific form:
\be
\h^\text{qp}_{\epsilon}[\R, \lambda] = \sum_{a, b = 1}^{B} \left(\R_{a\uparrow} \epsilon \R^*_{b\uparrow} +\lambda_{ab} \right) \fc_{a\uparrow} \fa_{b\uparrow} \, .
\ee
The self-energy, which can be derived from the Dyson equation~\cite{gga_lanata2017}, is independent of $\epsilon$ due to the locality of the self-energy.
Here we consider the analytical expression derived in Ref.~\cite{Lee2023AccuracyGRISB}, which resembles the pole-expansion form of the self-energy proposed in DMFT~\cite{Lee2023AccuracyGRISB, Savrasov2006ManyBodyES}.: 
\be
\begin{aligned}
\Sigma_{\uparrow}(\omega) &= \omega + i0^+ - \epsilon -G_{\epsilon\uparrow}^{-1}(\omega) \\
&= \omega[1-(\R_\uparrow^\dag R_\uparrow)^{-1}] + [R_{0}]^{-1} \lambda_0 [R_{0}^\dag]^{-1} \\
&\quad + \sum_{c=2}^{B} \frac{R_{0}^{-1} [\lambda_1]_{\uparrow c} [\lambda_1^\dag]_{c \uparrow} [R_{0}^\dag]^{-1} }{\omega + i 0^+ - [\lambda_2]_{cc}}\,. \label{eq: self_energy}
\end{aligned}
\ee
Here, the \(1 \times 1\) matrix \(R_0\) is related to the \(B \times 1\) matrix \(\R_{\uparrow}\) through a gauge transformation \(\R_{\uparrow} = u [R_0 \quad 0]^T\). Under this transformation, the \(B \times B\) spin-up block of \(\lambda\) is transformed as \(u^\dagger \lambda u = \begin{bmatrix} \lambda_0 & \lambda_1 \\ \lambda_1^\dagger & \lambda_2 \end{bmatrix}\), where \(\lambda_0\) is a \(1 \times 1\) submatrix and \(\lambda_2\) is diagonal. It is evident that the ghost orbitals introduce poles in the self-energy at the energy levels corresponding to the diagonal elements of \(\lambda_2\), with residues tied to the coupling \(\lambda_1\) between the ghost modes and the physical orbital.

Finally, one can obtain the quasi-particle weight as:
\be
\begin{aligned}
    Z &= [1-\partial\Re{\Sigma_{\uparrow}(\omega)}/\partial \omega|_{\omega\to 0} ]^{-1} \\
    &= \left[(\R_\uparrow^\dag R_\uparrow)^{-1} +  \sum_{c=2}^{B} \frac{R_{0}^{-1} [\lambda_1]_{\uparrow c} [\lambda_1^\dag]_{c \uparrow} [R_{0}^\dag]^{-1} }{[\lambda_2]_{cc}^2}\right]^{-1}. \label{eq: Z}
\end{aligned}
\ee
This can also be conveniently evaluated using numerical techniques such as automatic differentiation~\cite{jax2018github}.

\section{Adaptive Variational Quantum Imaginary-Time Evolution Algorithm}
The AVQITE algorithm is developed by integrating the strengths of two existing approaches: Variational Quantum Imaginary Time Evolution (VQITE)~\cite{VQITE} and Quantum Imaginary Time Evolution (QITE)~\cite{qite_chan20}. VQITE, equivalent to the quantum natural gradient minimization approach~\cite{stokes2020quantum}, is known for its efficient representation of quantum states using compact parameterized circuits, making it suitable for near-term quantum hardware with limited coherence time and gate fidelity. However, its accuracy depends on the expressiveness of the chosen ansatz, which may limit its ability to closely approximate the exact imaginary time evolution.

Meanwhile, QITE provides a systematic approach to evolving a quantum state along the imaginary time trajectory, ensuring convergence to the ground state of a given Hamiltonian, provided the initial state has a finite overlap with it. By dynamically expanding the evolution operator at each time step, similar to Trotterized circuits, QITE can closely approximate the exact imaginary time evolution as long as a sufficiently large operator support is available to construct the unitaries approximating the imaginary time propagator. However, in its standard formulation QITE leads to increasingly deep circuits as time progresses, with the circuit depth growing exponentially with the correlation domain size associated with the action of Hamiltonian terms. This rapid increase in circuit complexity poses a significant challenge for near-term quantum hardware, limiting its practical implementation.

AVQITE bridges the strengths of VQITE and QITE by dynamically adapting the variational ansatz during the imaginary time evolution process. A key observation is that the McLachlan variational principle, which underlies VQITE, naturally defines a McLachlan distance that quantifies the deviation between the trajectory of the variational state and the exact imaginary time evolution. This McLachlan distance serves as a measure of accuracy, providing a criterion for ansatz expansion.
To ensure the variational evolution, similarly to QITE, remains a close approximation to the exact imaginary time dynamics, AVQITE keeps the McLachlan distance below a preset threshold. This is achieved by expanding the ansatz only when necessary through the introduction of new unitaries to the circuit. The additional unitaries are chosen based on their effectiveness in reducing the McLachlan distance, with their generators selected from a predefined operator pool, an approach inspired by adaptive variational quantum eigensolvers~\cite{grimsleyAdaptiveVariationalAlgorithm2019, MayhallQubitAVQE}. 
This adaptive mechanism enables AVQITE to maintain compact circuits when the existing ansatz adequately captures the imaginary time evolution, ensure accuracy by selectively expanding the ansatz only when deviations become significant, and optimize computational efficiency by controlling circuit growth. This approach substantially reduces circuit depth compared to QITE~\cite{AVQITE, smqite, Alipanah2025QuantumQS}, making AVQITE more practical for near-term quantum hardware.

The variational ansatz in AVQITE is structured in a pseudo-Trotter form, where the evolution operator is expressed as a product of parameterized unitaries acting on a reference state \(\ket{\Psi_0} \). Specifically, 
the ansatz takes the form:
\be
\left| \Psi(\boldsymbol{\theta}) \right\rangle = 
\prod_{k} e^{-i \theta_k G_k} \left| \Psi_0 \right\rangle, \label{eq: ansatz}
\ee
where \( G_k \) are the generators of the unitary evolution, selected from a predefined operator pool \(\mathcal{P} \) based on their effectiveness in reducing the McLachlan distance. 
In all calculations we adopt the Jordan-Wigner encoding to translate the fermionic 
embedding Hamiltonian~\eqref{eq: heb1} into a qubit representation, expressed as a sum of weighted 
Pauli strings: \(\h^\mathrm{eb}[\D, \lambda^\mathrm{c}] = \sum_j w_j P_j \). Here \(w_j\) is a function of \(\D\) and \(\lambda^\mathrm{c}\), and the Pauli string \(P_j \in \{I, X, Y, Z \}^{\otimes N}\) is defined as a direct product of Pauli operators (\(X, Y, Z\)) plus the identity (\(I\)) over \(N=2(1+B)\) qubits. 
The reference state \( \ket{\Psi_0} \) is set as a classical product state, where the physical site and \( (B-1)/2 \) bath sites are fully occupied~\cite{mukherjee2023comparative}. Throughout the calculations we employ the qubit excitation operator pool \( \mathcal{P} \):
\be
\mathcal{P} = \{Y_i X_j \}_{i\neq j=1}^{N} \cup \{Y_i X_j X_k X_l, Y_i Y_j Y_k X_l\}_{i\neq j\neq k \neq l =1}^{N}, \label{eq: pool}
\ee
which consists of a set of Pauli strings of weight 2 and 4 with odd number of Pauli $Y$'s, including all possible permutations. This pool consists of individual qubit operators from the qubit-excitation pool and is directly related to the generators in the unitary coupled cluster ansatz with single and double excitations.
During the expansion of the AVQITE ansatz~\eqref{eq: ansatz} by appending unitaries with generators selected from the operator pool \( \mathcal{P} \), we prioritize generators that act on disjoint qubits. This strategy has been shown to produce compact, layer-wise dense circuits with reduced gate count~\cite{Zhang2025AdaptiveVQ, Anastasiou2024TETRIS}.

\begin{figure}[ht!]
	\centering
	\includegraphics[width=\linewidth]{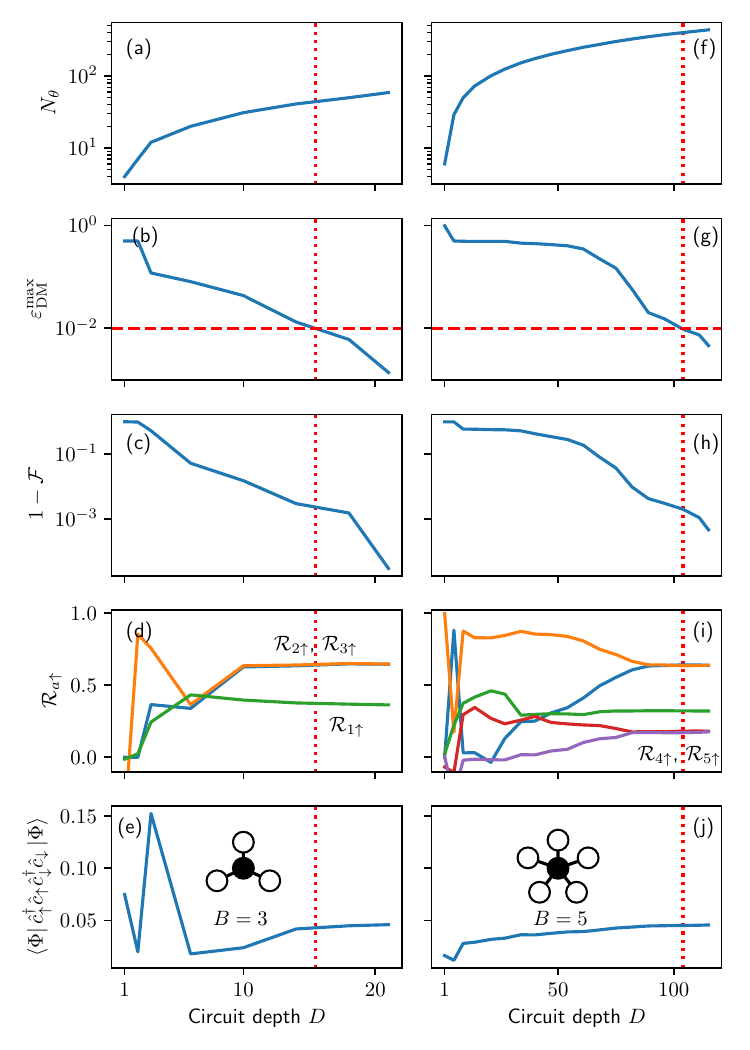}
	\caption{
	\textbf{Ground state preparation for the embedding Hamiltonians using AVQITE statevector simulations.} (a) The number of variational parameters $N_{\bth}$, as a function of circuit depth, $D$, for the embedding Hamiltonian with $B=3$, which amounts to 8 qubits. (b) The maximal error in the single-particle density matrix elements $\rho_{\mu\nu}$~\eqref{eq: 1pdm}, $\varepsilon_\mathrm{DM}^\mathrm{max} = \max_{\mu,\nu} \abs{\rho_{\mu\nu}[\bth] - \rho_{\mu\nu}^\mathrm{ED}}$. (c) The state infidelity, $1-\mathcal{F} = 1- \abs{\ovij{\Phi[\bth]}{\Phi^{\mathrm{ED}}}}^2$. (d) The renormalization factor $\R_{a\uparrow, \uparrow}$~\eqref{eq: R}. (e) The onsite double occupancy $\avihi{\Phi}{\cc_{\uparrow}\ca_{\uparrow}\cc_{\downarrow}\ca_{\downarrow}}$. (f-j) The same as (a-e) but for the embedding Hamiltonian with $B=5$, which amounts to 12 qubits. Superscript $\mathrm{ED}$ represents quantities obtained via exact diagonalization, while parameters $\bth$ refer to those evaluated from the adaptively generated parameterized ansatz. The horizontal dashed lines denote the accuracy requirement of $\varepsilon_\mathrm{DM}$ for typical GQCE(g) calculations, and the vertical dotted lines indicate the associated values for the other quantities of interest in the other panels.
	}
	\label{fig: B35}
\end{figure}

\section{Results}

\subsection{Statevector Simulations}
We first discuss the results of ground state preparation using AVQITE for the embedding impurity models~\eqref{eq: heb1} with $U=2.5$ and $B=3$ and $5$. The Hamiltonian parameters are obtained by performing self-consistent gGA calculations for the lattice model~\eqref{eq: sbhm} using classical algorithms, specifically exact diagonalization (ED) to solve for the ground state of the embedding impurity model~\eqref{eq: heb1}. Hubbard interaction $U=2.5$ sets the model at the boundary between the metallic phase and the Mott insulating phase with both solutions coexisting, which is a regime within the Hamiltonian parameter space that poses the greatest challenge for numerical simulations. We choose $B=3$ and $5$ to demonstrate the convergence of gGA calculations of the lattice model~\eqref{eq: sbhm} to the exact ground state solution. To showcase the ground state preparation of the impurity model~\eqref{eq: heb1}, we first simulate the AVQITE approach using a statevector simulator on a classical computer, which is free of hardware errors or sampling noise.

Figure~\ref{fig: B35} shows the AVQITE simulation results with increasing circuit depth $D$ for the two impurity models. The circuit depth $D$ is defined as the number of layers of unitary operations, where each layer consists of unitaries acting on mutually disjoint qubits. For calculations of the $B=3$ model, as $D$ increases from 1 to 21, the number of variational parameters $N_{\bth}$ grows from 4 to 59; meanwhile, the error in the density matrix $\varepsilon_\mathrm{DM}^\mathrm{max}$ reduces from 0.5 to 0.001 and the infidelity $1-\mathcal{F}$ drops from 0.997 to $3 \times 10^{-5}$, as shown in Fig.~\figref{fig: B35}{(a-c)}. Here $\varepsilon_\mathrm{DM}^\mathrm{max} = \max_{\mu,\nu} \abs{\rho_{\mu\nu}[\bth] - \rho_{\mu\nu}^\mathrm{ED}}$ is defined as the maximal error in the density matrix elements $\rho_{\mu\nu}[\bth]$ compared with the exact diagonalization results. In typical GQCE(g) calculations, we keep $\varepsilon_\mathrm{DM}$ below $0.01$. This error threshold is indicated by the horizontal dashed line in Fig.~\figref{fig: B35}{(c)}, and it corresponds to the infidelity of $1-\mathcal{F} = 2\times 10^{-3}$ and $D = 16$ layers of unitaries with $N_{\bth} = 44$ rotation angles. Figure~\ref{fig: B35}(d) shows that the three renormalization factors, $\R_{1\uparrow}$ and the degenerate $\R_{2\uparrow}$ and $\R_{3\uparrow}$, converge well with the criterion of $\varepsilon_\mathrm{DM} = 0.01$. Note that the degeneracy between $\R_{2\uparrow}$ and $\R_{3\uparrow}$ is broken at the shallower circuit depths ($D < 10$), indicating that the adaptive ansatz does not always preserve the model's symmetry. However, this artificial symmetry breaking can also serve as a useful metric to assess the convergence of AVQITE calculations. In Fig.~\ref{fig: B35}(e) we plot the convergence behavior of the onsite double occupancy $\avihi{\Phi}{\cc_{\uparrow}\ca_{\uparrow}\cc_{\downarrow}\ca_{\downarrow}}$, which reaches $0.046$ at large $D$ 
 as the calculation converges to the ED result.

When the model size increases from $B=3$ to $B=5$ (which corresponds to the increase in the number of qubits from $8$ to $12$), we observe a substantial growth of the circuit depth in the AVQITE calculations, as shown in Fig.~\figref{fig: B35}{(f-j)}. At the threshold $\varepsilon_\mathrm{DM} = 0.01$ the circuit reaches $D=104$ layers of unitaries with $N_{\bth} = 404$ parameters, while the infidelity remains small at $1-\mathcal{F} = 2\times 10^{-3}$. Two additional degenerate elements, $\R_{4\uparrow}$ and $\R_{5\uparrow}$, are introduced due to the increase in $B$. All renormalization factors converge well, reflecting the recovery of the symmetry for the degenerate pairs. At the same time, the double occupancy $\avihi{\Phi}{\cc_{\uparrow}\ca_{\uparrow}\cc_{\downarrow}\ca_{\downarrow}}$ reaches the same value of $0.046$ as for $B=3$, indicating its good convergence in the simulations of the lattice model~\eqref{eq: sbhm}. Another important physical quantity is the quasi-particle weight $Z$~\eqref{eq: Z}. In contrast to the double occupancy, we find it decreases from $0.13$ for $B=3$ to $0.10$ for $B=5$, implying slower convergence with $B$.

Table~\ref{tb: gqce} summarizes the results of GQCE(g) calculations for the Hubbard model on the Bethe lattice with bath orbital parameters $B = 3$ and 5, and compares them to the numerically exact DMFT solution using numerical renormalization group (NRG) as the impurity solver~\cite{Zitko2009NRG, zitko_2021_NRG, TRIQS2015}. For completeness, the result for $B = 1$, corresponding to GQCE with the conventional GA, is also included. The data confirm the effectiveness of GQCE(g) for $B > 1$, which significantly improves upon the $B = 1$ case and shows clear convergence to the DMFT+NRG values at $B = 5$ in terms of both double occupancy and quasiparticle weight. Meanwhile, the AVQITE circuit used for ground-state preparation of the embedding impurity model exhibits a steep increase in complexity: the number of variational parameters grows by a factor of 15 when increasing $B$ from 1 to 3, and by another factor of 9.2 from $B = 3$ to 5. The corresponding circuit depth increases by factors of 8 and 6.5, respectively.

\begin{table}[ht!]
\centering
\caption{Summary of GQCE(g) calculations for the bath orbital parameter $B=1$, 3, and 5, and their convergence toward the numerically exact solution of the Hubbard model on the Bethe lattice, as calculated using the DMFT+NRG approach~\cite{Zitko2009NRG, zitko_2021_NRG, TRIQS2015}.
}
\begin{tabular}{lcccc}
\toprule
 & $B=1$ & $B=3$ & $B=5$ & DMFT+NRG \\
\midrule
Double occupancy & 0.066 & 0.046 & 0.046 & 0.047 \\
Quasiparticle weight $Z$ & 0.45 & 0.13 & 0.10 & 0.09 \\
Number of angles $N_{\bth}$ & 3 & 44 & 404 & -- \\
Circuit depth $D$ & 2 & 16 & 104 & -- \\
Number of qubits & 4 & 8 & 12 & -- \\
\bottomrule
\end{tabular}
\label{tb: gqce}
\end{table}

\begin{figure}[ht!]
	\centering
	\includegraphics[width=\linewidth]{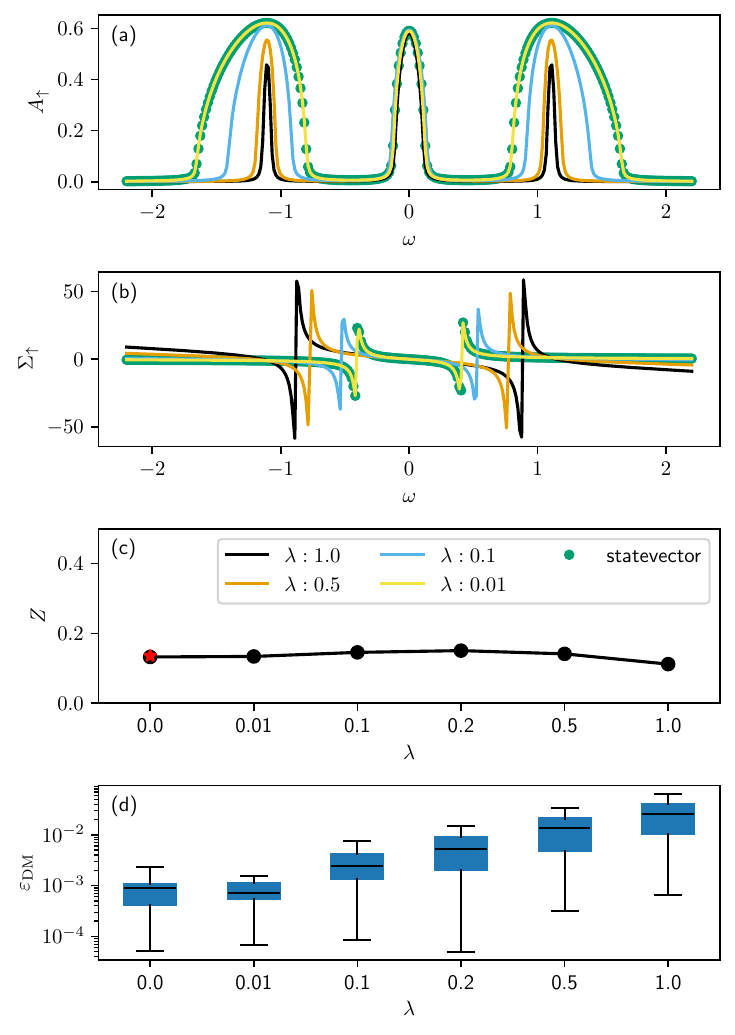}
	\caption{
    \textbf{Dependence of the simulated electronic structure of the Hubbard model on noise strength.} 
    (a) Evolution of the spectral function $A_{\uparrow}(\omega)$ with respect to the noise strength scaling factor $\lambda$ (see legend in panel c). 
    (b) Corresponding variations in the self-energy $\Sigma_{\uparrow}(\omega)$. 
    (c) Quasiparticle weight $Z$ as a function of $\lambda$, where the red star marks the noiseless $Z$ value. 
    (d) Effect of $\lambda$ on the error distribution of the density matrix, $\varepsilon_\mathrm{DM}$, relative to the statevector simulator result obtained in the absence of noise. The blue box plot represents the first, second (median), and third quartiles, while whiskers extend to the minimum and maximum values.
    The circuit used here has a depth of $D = 18$, as studied in Fig.~\figref{fig: B35}{(a-e)} with $B=3$ (8 qubits), and comprises 108 two-qubit gates after transpilation under the assumption of all-to-all qubit connectivity. We apply $10^5$ shots for the measurement of each observable.
	}
	\label{fig: spec}
\end{figure}

\subsection{Noisy Simulations}
The impact of noise~\cite{Khindanov2025} on AVQITE calculations has been investigated using a phenomenological noise model incorporating amplitude damping and dephasing channels~\cite{getelina2023adaptive}. In this work, we examine its impact on the estimation of the final density matrix and the subsequent determination of the spectral function and the self-energy. To this end, we adopt the noise model from Ref.~\cite{self2024protecting}, which includes state preparation and measurement (SPAM) errors along with depolarizing errors. Specifically, SPAM errors are modeled using a bit-flip channel applied after the qubit initialization (with the error rate $p_\mathrm{bi}$) and before any measurement in the circuit (with the error rate $p_\mathrm{bm}$): $\mathcal{E}_\mathrm{bi/bm}(\varrho) = (1-p_\mathrm{bi/bm})\varrho + p_\mathrm{bi/bm}X\varrho X$. 
At the same time, depolarizing error channel is applied after each single-qubit (1q) and two-qubit (2q) gate, with its operator-sum representations given by: $\mathcal{E}_\mathrm{1q}(\varrho) = (1-p_\mathrm{1q})\varrho + \frac{p_\mathrm{1q}}{3}(X\varrho X + Y\varrho Y + Z\varrho Z)$ and $\mathcal{E}_\mathrm{2q}(\varrho) = (1-p_\mathrm{2q})\varrho + \frac{p_\mathrm{2q}}{15}\sum_{P_2} P_2\varrho P_2$. Here $X, Y, Z$ refer to the single-qubit Pauli gates, while $P_2$ denotes the $15$ two-qubit Pauli gates $\{ I, X, Y, Z\}^{\otimes2}$ excluding the identity $II$. 
Following Ref.~\cite{self2024protecting} we set $p_\mathrm{bi} = p_\mathrm{1q} = 4\times 10^{-4}$ and $p_\mathrm{2q} = p_\mathrm{bm} = 3\times 10^{-3}$, according to the randomized benchmarking of the Quantinuum H1-2 hardware.
To analyze the sensitivity of the calculation results to noise, we introduce a noise strength scaling factor $\lambda$, which uniformly scales all error rates in the noise model~\cite{self2024protecting}.

Figure~\ref{fig: spec} presents the analysis of the impact of noise on the simulated electronic structure of the Hubbard model. As shown in Fig.~\figref{fig: spec}{(a)}, the spectral function \( A_{\uparrow}(\omega) \) exhibits a central resonant peak at the Fermi level, \( E_f = 0 \), along with the lower and upper Hubbard bands at $\omega \approx \pm U/2$, consistent with expectations for the Hubbard model at an intermediate interaction strength of \( U = 2.5 \). The shape of the central resonant peak is robust against noise, whereas the spectral weight of the Hubbard bands is significantly affected. For error rates comparable to those of the Quantinuum H1-2 device (\(\lambda=1\), black line), the middle resonant peak closely matches the exact result (blue circles), while the Hubbard bands are almost completely suppressed by noise. As \( \lambda \) decreases, the spectral weight of the Hubbard bands gradually recovers. For \(\lambda=0.01\) (yellow line), the Hubbard bands align with the exact result, indicating that the reduction in spectral weight at finite noise levels is primarily due to the gate and measurement errors rather than the statistical noise from the finite sampling with \( 10^5 \) shots.

Figure~\figref{fig: spec}{(b)} presents the effect of noise on the self-energy \( \Sigma_{\uparrow}(\omega) \). Consistent with the noise dependence of the spectral function shown in Fig.~\figref{fig: spec}{(a)}, the dip and the peak in \( \Sigma_{\uparrow}(\omega) \) associated with the lower and upper Hubbard bands exhibit significant shifts as the error rates decrease. In contrast, the central region of \( \Sigma_{\uparrow}(\omega) \) around \( \omega=0 \) remains largely unaffected by noise, mirroring the robustness of the central resonant peak in the spectral function. Consequently, the quasiparticle weight \( Z \), determined by the slope of \( \Sigma_{\uparrow}(\omega) \) at \( \omega=0 \), varies within the narrow interval \( [0.08, 0.16] \) as shown in Fig.~\figref{fig: spec}{(c)}.

In the above analysis noise effects manifest through errors in the density matrix, \( \varepsilon_\mathrm{DM} \), quantified by the element-wise absolute differences between the simulated noisy one-particle density matrix \( \rho_{\mu\nu} \) and the exact values for the given circuit, along with the corresponding discrepancy in the double occupancy $\avihi{\Phi}{\cc_{\uparrow}\ca_{\uparrow}\cc_{\downarrow}\ca_{\downarrow}}$. The distribution of \( \varepsilon_\mathrm{DM} \) as a function of noise strength is presented in the box plot of Fig.~\figref{fig: spec}{(d)}. At noise scale \( \lambda=1 \) the maximal error reaches 0.11, while the median is 0.04. As \( \lambda \) decreases, \( \varepsilon_\mathrm{DM} \) and its spread generally reduces. At \( \lambda=0.01 \) the errors become dominated by sampling noise (\(\lambda = 0\)).

\subsection{Density Matrix Measurement with Error Detection}
The noisy simulation results presented in the previous subsection demonstrate that the error rates in current QPUs are substantial, implying the necessity to utilize quantum error correction (QEC) techniques to achieve accurate calculations. Given recent experimental advancements in QEC~\cite{krinner2022realizing, sivak2023real, bluvstein2024logical, paetznick2024demonstration, gupta2024encoding, AI2024QECBelow}, we aim to benchmark the impact of quantum error detection (QED)~\cite{Knill2004schemes,Knill2004threshold,gowrishankar2024logical} on the accuracy of density matrix measurements in our circuits.  
Although QED relies on post-selection to achieve error reduction, its implementation generally requires fewer physical qubits, simpler syndrome measurement circuits, and no decoding procedure compared to QEC, making it a more accessible technique for early fault-tolerant quantum computing.

\begin{figure}[ht!]
	\centering
	\includegraphics[width=\linewidth]{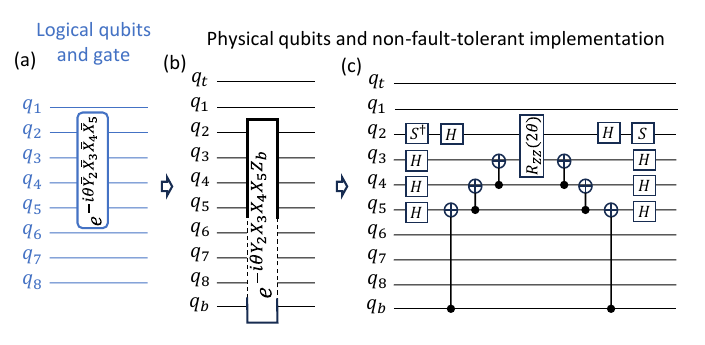}
	\caption{
	\textbf{Illustration of the non-fault-tolerant implementation of a logical Pauli evolution gate.} (a) An example of a logical Pauli evolution gate in the AVQITE ansatz for simulations of the impurity model with eight logical qubits. (b) Replacing the generator of the logical Pauli evolution unitary with the associated physical Pauli operators according to Eq.~\eqref{eq: XYgates}, in the physical qubit representation based on \( \llbracket 10, 8, 2\rrbracket \) Iceberg code. (c) The standard CNOT ladder decomposition with \( R_\mathrm{ZZ} \) gate.
	}
	\label{fig: gate}
\end{figure}

\begin{figure*}[ht!]
	\centering
	\includegraphics[width=\textwidth]{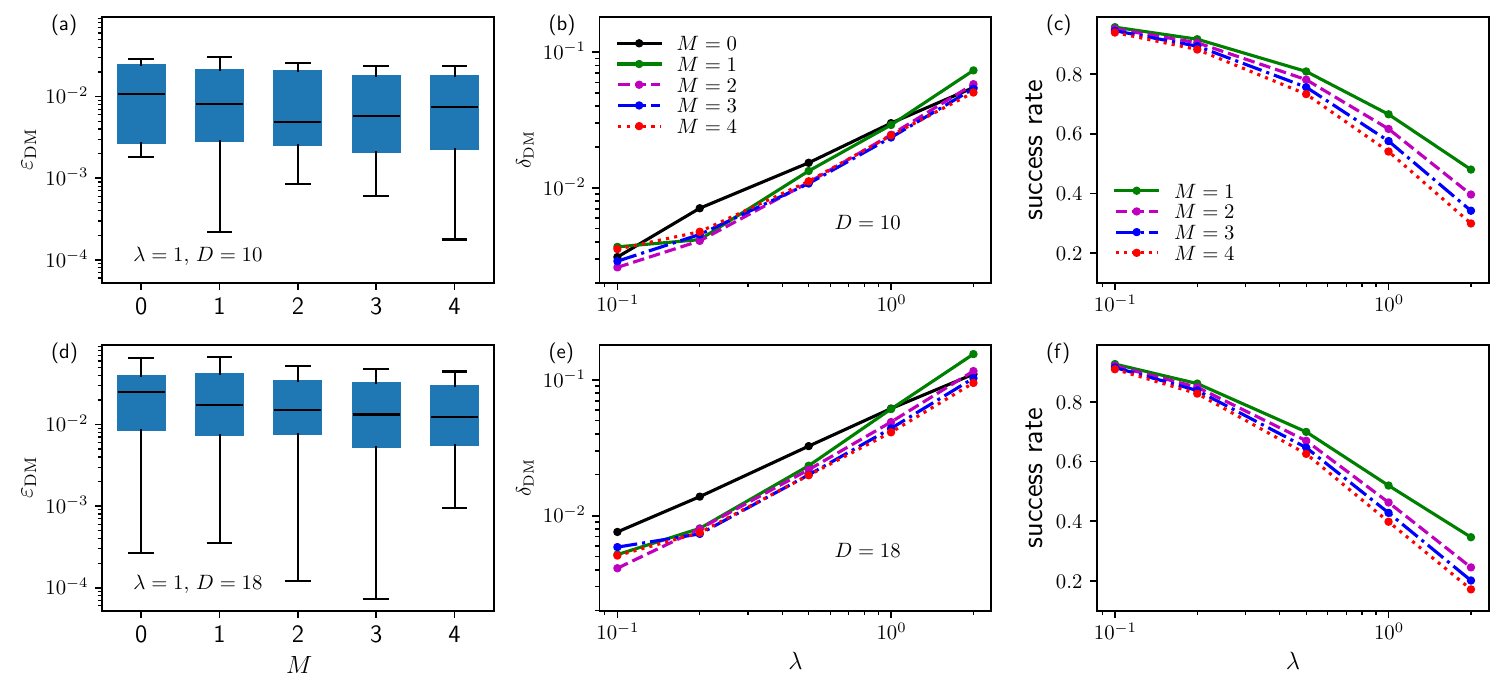}
	\caption{
    \textbf{Error reductions in the density matrix measurements with the help of the Iceberg QED code.}  
    (a) Box plot of the density matrix errors, \( \varepsilon_\mathrm{DM} \), as a function of the number of stabilizer syndrome measurements, \( M \), for an eight-qubit circuit with depth \( D = 10 \), encoded using the \( \llbracket 10, 8, 2 \rrbracket \) Iceberg code. The box plot displays the standard first, second (median), and third quartiles, with whiskers extending to the minimum and the maximum values. 
    (b) The corresponding trace distance between the measured one-particle density matrix \(\rho_\mathrm{m}\) and its exact value \(\rho_\mathrm{ref}\), \( \delta_{\mathrm{DM}} = \frac{1}{2}\left\Vert \rho_\mathrm{m} -\rho_\mathrm{ref}\right\Vert_\mathrm{F}\), plotted as a function of the noise scaling factor \( \lambda \) and \( M \). Here $\left\Vert .\right\Vert_\mathrm{F} $ denotes the Frobenius norm.
    (c) The success rate, defined as the ratio of shots without detected errors to the total number of shots, as a function of \( \lambda \) and \( M \).  
    (d, e, f) The same as (a, b, c), but for an eight-qubit circuit with depth \( D = 18 \). 
    The results for \( M = 0 \) are obtained from simulations using circuits without the Iceberg encoding, where we apply \( 10^5 \) shots for measuring each Pauli string in the observables. 
    For logical circuits, the number of shots is increased by a factor of up to 5, depending on \( M \), to ensure that the total number of successful samples remains \( 10^5 \).
	}
	\label{fig: dm_iceberg}
\end{figure*}

To encode \( k = 8 \) logical qubits required to represent the embedding Hamiltonian with \( B = 3 \), we adopt the \( \llbracket k+2, k, 2\rrbracket \) Iceberg QED code~\cite{self2024protecting, he2024performance}. The Iceberg code employs the top \( t \), the bottom \( b \), and \( k \) qubits in between for encoding, where \(k\) is even. Additionally, two ancilla qubits are required for syndrome measurements, resulting in a total of \( n=12 \) physical qubits in our circuit.
The code has two stabilizers:  
\(
S_{X} = X_{t}X_{b} \bigotimes_{j=1}^{k} X_{j},\ S_{Z} = Z_{t}Z_{b} \bigotimes_{j=1}^{k} Z_{j},
\)
and the single-qubit logical \(\Bar{X}_i \) and \(\Bar{Z}_i \) operators are defined on two physical qubits: 
\be
\Bar{X}_i = X_i X_t,\ \Bar{Z}_i = Z_i Z_b. \label{eq: XYgates}
\ee
The initialization to a logical all-zero state \( |\Bar{0}\rangle \) amounts to preparing the GHZ state on $(k+2)$ physical qubits.
As a distance-2 code, the Iceberg code can detect any single-qubit error. However, certain two-qubit errors, such as those induced by $X_i X_j$ in our noise model, keep the state within the code space but result in a logical error.
Within the Iceberg code the initialization, the syndrome measurements and the final measurement can be constructed in a fault-tolerant manner such that a failure of any single component would not create an undetectable logical error. The corresponding circuits for fault-tolerant implementations can be found in Ref.~\cite{self2024protecting}.

Although the fault-tolerant logical gate set for the Iceberg code, including the non-Clifford logical CCZ gate, is known~\cite{Chao2018FaultTQC}, directly implementing these logical circuits remains costly for near-term QPUs.
Therefore, to implement the logical circuit for the AVQITE ansatz~\eqref{eq: ansatz}, we adopt non-fault-tolerant logical unitary gates of the form \( e^{-i\theta_k \Bar{G}_k} \), where each \( \bar{G}_k \) is selected from the operator pool~\eqref{eq: pool}, as follows:  
\begin{align}
    e^{-i\theta\Bar{Y}_i \Bar{X}_j} &= e^{-i\theta Y_i X_j Z_b},\label{eq:log_to_phys_YX} \\
    e^{-i\theta\Bar{Y}_i \Bar{X}_j \Bar{X}_k \Bar{X}_l} &= e^{-i\theta Y_i X_j X_k X_l Z_b}, \\
    e^{-i\theta\Bar{Y}_i \Bar{Y}_j \Bar{Y}_k \Bar{X}_l} &= e^{-i\theta Y_i Y_j Y_k X_l Z_b}. \label{eq:log_to_phys_YYYX}
\end{align}
The physical multi-qubit unitaries are further decomposed using a CNOT ladder into single-qubit gates and the \( R_\mathrm{ZZ} \) gate, which is a native gate for trapped-ion devices. An example of this procedure is illustrated in Fig.~\ref{fig: gate}.

To evaluate the effectiveness of the QED code in reducing errors in GQCE(g) calculations, we focus on the measurement of the density matrix using circuits analyzed in Fig.~\figref{fig: B35}{(a-e)} with depths \( D = 10 \) and \( D = 18 \). When these circuits are unencoded and transpiled under the assumption of all-to-all qubit connectivity, the number of two-qubit gates in them is equal to \( N_{\rm 2q}=50 \) for \( D = 10 \) and \( N_{\rm 2q}=108 \) for \( D = 18 \). However, when encoded using the Iceberg code and the rules ~\eqref{eq:log_to_phys_YX}-\eqref{eq:log_to_phys_YYYX}, the number of physical two-qubit gates in these circuits increases to \( N_{\rm 2q}=130 \) for \( D = 10 \) and \( N_{\rm 2q}=239 \) for \( D = 18 \). 
Furthermore, the fault-tolerant implementation of syndrome measurements for both \( S_X \) and \( S_Z \) stabilizers in the Iceberg code introduces an additional \( N_{\rm 2q}=20 \) two-qubit gates per syndrome measurement. 

We analyze the measured density matrix error, \( \varepsilon_\mathrm{DM} \), as a function of the number of syndrome measurements \( M \), which are uniformly distributed throughout the circuit, with the final syndrome measurement placed immediately before the observable measurement.
If the logical observables of interest were Pauli strings composed solely of \(\bar{Z}_i\) operators, it would be sufficient to directly measure the physical qubits in the computational basis and infer both the logical observable values and the \( S_{Z} \) stabilizer outcomes after post-processing, as was done in Ref.~\cite{self2024protecting}.
However, the qubit representation of the observable in our case --- namely, the density matrix --- includes Pauli strings containing \(\bar{X}_i\) and \(\bar{Y}_i\) operators.
One way to perform the measurement in this case would be to apply the appropriate logical Hadamard \(\bar{H}\) and phase \(\bar{S}^\dag\) gates prior to the final measurement in the computational basis.
The final round of syndrome measurements then would be able to detect any single-qubit errors that may have occurred within the physical implementations of these logical gates.

Unfortunately,  both fault-tolerant and non-fault-tolerant implementations of the logical \(\bar{H}\) and \(\bar{S}^\dag\) gates would necessarily introduce additional overhead in the number of physical two-qubit gates, which could in turn lead to higher error rates given that our error model includes two-qubit errors.
To avoid this overhead, we instead perform the final measurement by first converting the logical observable Pauli strings into their physical representations, and then directly  applying the physical \( H \) and \( S^\dag \) gates to the relevant qubits prior to the measurement in the computational basis, without any post-processing step at the end. 
Although physical \( H \) and \( S^\dag \) gates within this approach would move the final circuit state outside of the code space --- thereby rendering any subsequent single-qubit errors undetectable --- the introduced error is expected to be minimal, as it involves only a small number of single-qubit physical gates with low error rates.

In Fig.~\figref{fig: dm_iceberg}{(a)}, we show the error distribution of \(\varepsilon_\mathrm{DM}\) as a function of the number of syndrome measurements \(M\) for the circuit with \(D=10\), obtained using noisy simulations at a noise scaling factor of \(\lambda=1\). Compared to the results for the unencoded circuit (\(M=0\)), the box plot demonstrates a noticeable improvement in \(\varepsilon_\mathrm{DM}\) for logical circuits with \(M\neq0\) syndrome measurements, particularly for the median when \(M=2\). 
To further quantify the effectiveness of using QED, we consider the trace distance \(\delta_{\mathrm{DM}} = \frac{1}{2} \left\Vert \rho_\mathrm{m} - \rho_\mathrm{ref} \right\Vert_\mathrm{F}\), which measures the Frobenius norm of the difference between the measured one-particle density matrix \(\rho_\mathrm{m}\) and the reference values \(\rho_\mathrm{ref}\) from the exact noiseless simulations. In Fig.~\figref{fig: dm_iceberg}{(b)}, we plot \(\delta_{\mathrm{DM}}\) as a function of the noise scaling factor \(\lambda\) for \(M \in [0, 4]\). Consistent with Fig.~\figref{fig: dm_iceberg}{(a)}, at \(\lambda=1\), \(\delta_{\mathrm{DM}}\) remains similar between \(M=0\) and \(M=1\). However, a noticeable decrease in \(\delta_{\mathrm{DM}}\) is observed for \(M=2\), while further increases in \(M\) (\(M=3,4\)) yield negligible additional improvements.  

The dependence of the observable measurement accuracy on \(M\) can be understood as a competition between two factors. On one hand, syndrome measurements and subsequent postselection reduce the total error by discarding samples with detected errors. On the other hand, the encoding and the insertion of syndrome measurements introduce additional overhead in the number of qubits and gates, thus increasing the total error due to the greater circuit depth and complexity. This trade-off naturally depends on the gate error rates, as illustrated in Fig.~\figref{fig: dm_iceberg}{(b)}. For the larger noise scaling factor, \(\lambda=2\), the value of \(\delta_{\mathrm{DM}}\) is the highest for \(M=1\), while the values for other \(M\) are close to each other. As \(\lambda\) decreases, values of \(\delta_{\mathrm{DM}}\) for finite \(M > 0\) become consistently smaller than those for \(M=0\), and reach a similar level for all \(M > 0\) at \(\lambda=0.2\), where the error is reduced by 40\% compared to the unencoded case. Further reducing the noise to \(\lambda=0.1\) causes the results for all \(M\) to merge, with the result at an intermediate \(M=2\) achieving the lowest value.
Figure~\figref{fig: dm_iceberg}{(c)} shows the success rates, defined as the fraction of samples without detected errors, which decrease with increasing \(\lambda\) and \(M\), as expected. The minimum success rate reaches approximately 30\% for \(M=4\) at \(\lambda=2\). In all QED simulations presented in this work, we gradually increase the number of shots for the observable measurements to compensate for the decreasing success rate, ensuring that the sample size for the expectation value estimation remains constant at \(10^5\). This approach eliminates the varying sample size effects in the error analysis.

The results for the deeper circuit with \(D=18\) are presented in Fig.~\figref{fig: dm_iceberg}(d)-(f). While \(\varepsilon_\mathrm{DM}\) is generally larger and has a wider distribution than in the case of a shallower circuit considered above, it exhibits a more systematic trend of error reduction with increasing \(M\), as can be observed in Fig.~\figref{fig: dm_iceberg}{(d)}. This trend is consistent with the behavior of \(\delta_\mathrm{DM}\) at \(\lambda=1\) illustrated in Fig.~\figref{fig: dm_iceberg}{(e)}.  
Here, similarly to Fig.~\figref{fig: dm_iceberg}{(b)}, a reduction in \(\delta_\mathrm{DM}\) for finite \(M > 0\) is observed as \(\lambda\) decreases. This reduction, however, is more pronounced in the case of the deeper circuit.
Again, as in the case of Fig.~\figref{fig: dm_iceberg}{(b)}, the values of \(\delta_\mathrm{DM}\) converge at \(\lambda=0.2\) and reach a minimum for the intermediate number of syndrome measurements, \(M=2\), at \(\lambda=0.1\). The success rates, shown in Fig.~\figref{fig: dm_iceberg}{(f)}, expectedly decrease with increasing \(\lambda\) and \(M\), reaching approximately 20\% for \(M=4\) at \(\lambda=2\).  

\begin{figure}[ht!]
	\centering
	\includegraphics[width=\linewidth]{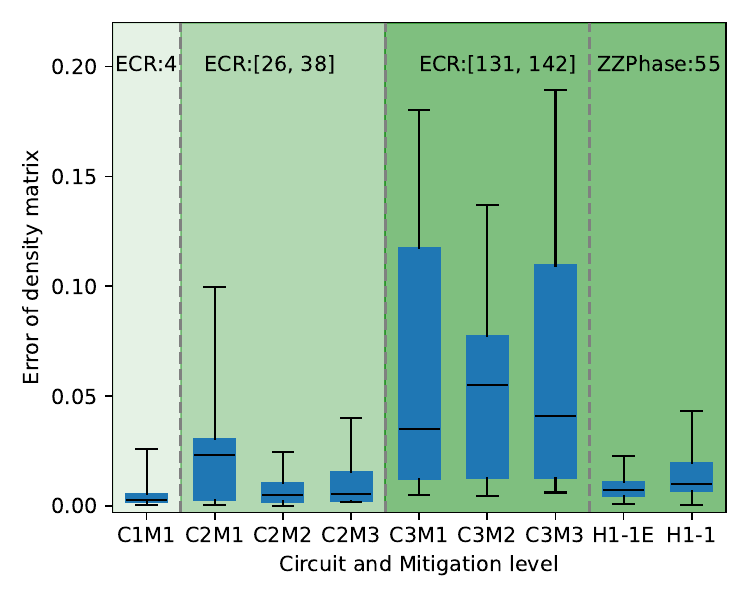}
	\caption{
	\textbf{Error analysis of the density matrix measured on QPUs for circuits across different depths.} Box plot for \(\varepsilon_\mathrm{DM}\), the errors in the measured one-particle density matrix~\eqref{eq: 1pdm} and the double occupancy, $\avihi{\Phi}{\cc_{\uparrow}\ca_{\uparrow}\cc_{\downarrow}\ca_{\downarrow}}$, with circuits of depth $D=1$(C1), $3$(C2) and $10$(C3) on IBM QPUs. 
    Results for circuit C3 obtained using Quantinuum’s H1-1 QPU and its noise-modeled emulator (H1-1E) without any error mitigation are also presented.
    The box plot represents the standard first, second (median), and third quartiles, with whiskers indicating the minimum and maximum values. 
    Note that due to the limited gate set and qubit connectivity in IBM QPUs, the actual circuit depth after transpilation increases to $12$ for C1, $61$ for C2, and $255$ for C3. The range of native two-qubit echoed cross-resonance (ECR) gates, resulting from stochastic transpilation, is also listed. 
    In contrast, the all-to-all qubit connectivity of the H1-1 device allows the transpiled C3 circuit to achieve a lower depth of 76, using 55 two-qubit ZZ-phase gates.
    Three levels of error mitigation (M1, M2 and M3) as implemented in IBM circuit function in Qiskit are investigated. 
    We set a target precision of $0.01$ for the density matrix element measurement, requiring $10^4$ shots per measurement circuit.
    For each CM group, we repeat the density matrix evaluations three times, selecting the least-busy available IBM QPUs for each run. The QPUs used include ibm\_brisbane, ibm\_brussels, ibm\_strasbourg, and ibm\_torino across different executions. 
    For calculations on Quantinuum's backends, we use $10^3$ shots per measurement circuit due to limited quantum resource budget. The experiments on H1-1 QPU were run on May 22, 2025.
    The error of each density matrix element is computed as the absolute difference between the average value over three runs within each CM group (except for H1-1 calculations, which use a single run) and the exact result obtained from the statevector representation of the circuit, as discussed in Fig.~\ref{fig: B35}. 
    The green backgrounds highlight the three circuits, C1–C3, arranged in order of increasing depth.
	}
	\label{fig: dm_ibm}
\end{figure}

\subsection{Density Matrix Measurement on QPUs}

Direct execution of AVQITE calculations on QPUs is challenging due to a significant overhead in the number of measurements. This overhead can potentially be alleviated by incorporating classical methods for partial evaluation of the quantum geometric tensor~\cite{Zhang2025AdaptiveVQ}. As an essential step towards executing AVQITE on QPUs at scale, we examine the measurement of the one-particle density matrix, arising in GQCE(g) calculations, on IBM QPUs for circuits with increasing depth. Our goal is to assess the impact of native gate constraints, qubit connectivity limitations, hardware errors, and the effectiveness of error mitigation techniques in specific applications.

The box plot in Fig.~\ref{fig: dm_ibm} presents the error analysis of the measured density matrix on IBM QPUs for circuits of varying depths. The $x$-axis represents different circuit depths (C1, C2, and C3) combined with different levels of error mitigation (M1, M2, and M3). The $y$-axis quantifies the error in the density matrix. From C1 to C3, the numbers of echoed cross-resonance (ECR) gates increases significantly, as shown in Fig.~\ref{fig: dm_ibm}. 
As implemented in the IBM circuit function~\cite{Qiskit}, M1 combines dynamical decoupling, measurement twirling, and twirled readout error extinction techniques to mitigate errors in quantum circuits and measurements. M2 builds on M1 by incorporating gate twirling and zero-noise extrapolation (ZNE) via digital gate folding. M3 further refines ZNE by replacing gate folding with Probabilistic Error Amplification (PEA), where noise scaling is applied based on the twirled noise model learned for each layer of entangling gates.

As clearly illustrated in Fig.~\ref{fig: dm_ibm}, circuits with larger depths exhibit a higher density matrix error, with a notable rise from C1 to C3. For the simplest circuit (C1), the errors are already low ($\sim 0.01$) at the lowest mitigation level M1, suggesting low errors in executing short-depth circuits on QPUs and the effectiveness of error mitigation. However, as the circuit depth increases (C2 and C3), the errors escalate, particularly in C3. Here the third quartile rises significantly, and the whiskers extend to much higher values, demonstrating greater variability in performance depending on the circuit complexities.

The impact of error mitigation techniques (M1, M2, and M3) is particularly evident for circuit C2, as shown in Fig.~\ref{fig: dm_ibm}. Mitigation levels M2 and M3 significantly reduce errors compared to M1, bringing C2 closer to the accuracy of C1 and highlighting the effectiveness of ZNE. However, for the deeper circuit C3, error variations remain large across all mitigation levels, suggesting that hardware noise and gate infidelities are the primary factors limiting accuracy. This calls for further investigation into more effective error reduction strategies. 
In our circuit calculations on QPUs, we also observe that ZNE with digital folding in M2 performs slightly better than ZNE with PEA in M3.

The last two datasets in Fig.~\ref{fig: dm_ibm} show results for circuit C3 measured on Quantinuum’s H1-1 QPU and its noisy emulator, H1-1E, both without error mitigation. The observed errors are significantly smaller than those from IBM QPUs (C3M1–M3), primarily due to the simpler transpiled circuit enabled by all-to-all qubit connectivity, as well as higher gate fidelities and lower readout errors. Notably, the errors for circuit C3 on H1-1 are comparable to those for circuit C2 on IBM QPUs (C2M2, C2M3), reflecting their closer circuit depths and two-qubit gate counts after transpilation on the respective backends.

\section{Conclusion and Outlook}
In conclusion, we extend the Gutzwiller quantum-classical embedding framework by incorporating the ghost Gutzwiller approximation to improve both ground-state and spectral function calculations for strongly correlated materials. In this GQCE(g) approach, the simulation of bulk systems is mapped onto coupled eigenvalue problems of a noninteracting quasiparticle Hamiltonian, which can be efficiently solved on classical CPUs, and many-body impurity models, which require QPUs for scalable treatment. 

To enhance the accuracy of GQCE(g), one introduces a set of ghost orbitals, controlled by an integer parameter \( B \), as additional fermionic modes in the quasiparticle Hamiltonian. This extension systematically increases the variational degrees of freedom of the reference noninteracting wavefunction. Consequently, the embedding impurity model in GQCE(g) includes an additional \((B-1)\nu\) bath orbitals beyond the \(\nu\) bath orbitals in standard GQCE, increasing the model's complexity. 
Nevertheless, compared to DMFT with bath discretization, the classical implementation of gGA has demonstrated better convergence with respect to the number of bath orbitals for ground state property calculations~\cite{gga_lanata2017, Frank2021QuantumED, Lee2023AccuracyGRISB}. Moreover, gGA maintains higher computational efficiency by requiring only the ground state solution of the impurity model, which is significantly simpler than computing the frequency-resolved single-particle Green's function in DMFT.  

We demonstrate the GQCE(g) calculations on the single-band Hubbard model on a Bethe lattice, employing the adaptive variational quantum imaginary time evolution algorithm as the embedding impurity solver. We examine the convergence behavior for the most challenging parameter regime, when the system is at the phase boundary between the correlated metal and the Mott insulator. The AVQITE circuit depth is shown to increase by six-fold as \( B \) grows from 3 to 5.

To assess the impact of noise, we employ a noise model with parameters matching the Quantinuum H1-2 QPU and analyze its effects on the spectral function, self-energy, and other local quantities. Additionally, we investigate the sensitivity of these results to noise by uniformly scaling its strength. Our findings reveal that the central resonance peak of the spectral function remains robust against noise, whereas the side Hubbard bands exhibit noise dependence, converging well at the noise scaling factor of \( \lambda = 0.1 \) and fully converging at \( \lambda = 0.01 \).

Furthermore, we explore the effectiveness of the \( \llbracket 10, 8, 2 \rrbracket \) Iceberg QED code in reducing the errors in our embedding impurity model simulations. Through the analysis of the density matrix error distributions \(\varepsilon_\mathrm{DM}\) and trace distance \(\delta_\mathrm{DM}\), we showcase a sizable error reduction when employing the Iceberg code, compared to the unencoded circuit. The error reduction becomes most prominent when increasing the circuit depth \(D\) from 10 to 18 and reducing the noise strength by a scaling factor of \(\lambda=0.2\). 

Moreover, to assess the performance of current quantum hardware for GQCE(g) applications, we measure the density matrix on IBM superconducting transmon qubit-based QPUs across different circuit depths using various levels of error mitigation techniques. Due to native gate constraints and limited qubit connectivity, the circuit depth and number of two-qubit gates become substantially larger after transpilation. 
For instance, the circuit with \(D=10\) amounts to \(N_{\rm 2q}=50\) two-qubit gates assuming all-to-all qubit connectivity and the native \(R_\mathrm{ZZ}\) gate, and that number increases further to \(N_{\rm 2q}>130\) after the transpilation to IBM QPUs.
While basic error mitigation (M1) effectively suppresses errors for shallow circuits, more advanced techniques (M2 and M3) incorporating zero-noise extrapolation significantly improve the accuracy for intermediate-depth circuits. However, for the deeper circuit C3, the measured density matrix still exhibit substantial error and variability, suggesting the need for further improvements in the error reduction strategies. 
We demonstrate that the all-to-all qubit connectivity in alternative architectures, such as Quantinuum’s trapped-ion H1-1 QPU, enables a significant reduction in transpiled circuit depth and two-qubit gate count, leading to much smaller errors in expectation value estimations.

Looking ahead, several directions arise from our findings. First, while the converged AVQITE circuits demonstrate significant compression, their depths may still exceed the reliable execution capabilities of current QPUs. To address this, a quantum subspace expansion approach, such as combining a partially converged AVQITE ansatz with a maximally QPU-compatible circuit depth and Krylov-type subspace generation~\cite{Getelina2024QuantumSE}, may offer a potential path for leveraging quantum computing in GQCE(g) calculations. Incorporating advanced error mitigation techniques, quantum error detection and partial error correction is essential for successful applications in the pre-fault-tolerant and early-fault-tolerant quantum computing era. 

Second, another avenue worth exploring is quantum-selected configuration interaction~\cite{kanno2023quantum, robledo2024chemistry, Yu2025QuantumCA, Mikkelsen2025QuantumSCI}, where the ground state is obtained by classically diagonalizing the Hamiltonian in a subspace spanned by bit-strings sampled from QPUs with relatively shallow circuits, making the choice of an optimal QPU-compatible circuit a key consideration. Additionally, these approaches can be integrated with the recent observation that the effective subspace dimension of embedding impurity models remains small in self-consistent GQCE(g) calculations, as indicated by principal component analysis of gGA calculations for the Hubbard model on various lattices~\cite{Frank2024activeLA}. This insight may further enhance the feasibility and efficiency of GQCE(g) calculations.

Third, extending GQCE(g) calculations to multi-orbital models is a crucial step toward broader simulations of strongly correlated quantum materials. The classical implementation of gGA has demonstrated its applicability to \(d\)-electron systems such as NiO, using the density matrix renormalization group as an impurity solver~\cite{Lee2024ChargeSC}. However, extending these calculations to rare-earth or actinide materials, where \(f\)-electron correlations play a pivotal role, remains a formidable challenge. Therefore, it represents a natural opportunity for achieving practical quantum advantage in GQCE(g) simulations of \(f\)-electron systems.

Finally, extending GQCE(g) to nonequilibrium systems by incorporating time-dependent gGA will enable the study of out-of-equilibrium phenomena~\cite{schiro10tdga,Guerci2023TDgGA}, such as ultrafast electronic dynamics and transport properties. This approach could help explore transient behaviors and nonequilibrium steady states in strongly correlated materials, providing useful theoretical insights into their real-time response to external perturbations. Such developments may contribute to a better understanding of quantum materials at ultrafast timescales and their potential applications.

\section*{Acknowledgment}
We are grateful for discussions with Peter P. Orth, Gabriel Kotliar, Tsung-Han Lee, Davide Venturelli, Erik Gustafson, and M. Sohaib Alam.
This work was supported by the U.S. Department of Energy (DOE), Office of Science, Basic Energy Sciences, Materials Science and Engineering Division, including the grant of computer time at the National Energy Research Scientific Computing Center (NERSC) in Berkeley, California. The research was performed at the Ames National Laboratory, which is operated for the U.S. DOE by Iowa State University under Contract No. DE-AC02-07CH11358.
Calculations on quantum hardware were partially supported by the U.S. Department of Energy, Office of Science, National Quantum Information Science Research Centers, Co-design Center for Quantum Advantage (C2QA) under contract number DE-SC0012704 (Y.-X.Y.), and partially supported by IBM-HBCU Quantum Center and Southern University group (C.S. and H.M.B.). This work was supported in part by the U.S. Department of Energy, Office of Science, Office of Workforce Development for Teachers and Scientists (WDTS) under the Visiting Faculty (VFP) Program (H.M.B.) and the Science Undergraduate Laboratory Internships (SULI) program (C.S.).

\bibliography{ref}

\begin{thebibliography}{66}%
\makeatletter
\providecommand \@ifxundefined [1]{%
 \@ifx{#1\undefined}
}%
\providecommand \@ifnum [1]{%
 \ifnum #1\expandafter \@firstoftwo
 \else \expandafter \@secondoftwo
 \fi
}%
\providecommand \@ifx [1]{%
 \ifx #1\expandafter \@firstoftwo
 \else \expandafter \@secondoftwo
 \fi
}%
\providecommand \natexlab [1]{#1}%
\providecommand \enquote  [1]{``#1''}%
\providecommand \bibnamefont  [1]{#1}%
\providecommand \bibfnamefont [1]{#1}%
\providecommand \citenamefont [1]{#1}%
\providecommand \href@noop [0]{\@secondoftwo}%
\providecommand \href [0]{\begingroup \@sanitize@url \@href}%
\providecommand \@href[1]{\@@startlink{#1}\@@href}%
\providecommand \@@href[1]{\endgroup#1\@@endlink}%
\providecommand \@sanitize@url [0]{\catcode `\\12\catcode `\$12\catcode `\&12\catcode `\#12\catcode `\^12\catcode `\_12\catcode `\%12\relax}%
\providecommand \@@startlink[1]{}%
\providecommand \@@endlink[0]{}%
\providecommand \url  [0]{\begingroup\@sanitize@url \@url }%
\providecommand \@url [1]{\endgroup\@href {#1}{\urlprefix }}%
\providecommand \urlprefix  [0]{URL }%
\providecommand \Eprint [0]{\href }%
\providecommand \doibase [0]{https://doi.org/}%
\providecommand \selectlanguage [0]{\@gobble}%
\providecommand \bibinfo  [0]{\@secondoftwo}%
\providecommand \bibfield  [0]{\@secondoftwo}%
\providecommand \translation [1]{[#1]}%
\providecommand \BibitemOpen [0]{}%
\providecommand \bibitemStop [0]{}%
\providecommand \bibitemNoStop [0]{.\EOS\space}%
\providecommand \EOS [0]{\spacefactor3000\relax}%
\providecommand \BibitemShut  [1]{\csname bibitem#1\endcsname}%
\let\auto@bib@innerbib\@empty
\bibitem [{\citenamefont {Ivanov}\ \emph {et~al.}(2023)\citenamefont {Ivanov}, \citenamefont {S\"underhauf}, \citenamefont {Holzmann}, \citenamefont {Ellaby}, \citenamefont {Kerber}, \citenamefont {Jones},\ and\ \citenamefont {Camps}}]{Ivanov2023QuantumCP}%
  \BibitemOpen
  \bibfield  {author} {\bibinfo {author} {\bibfnamefont {A.~V.}\ \bibnamefont {Ivanov}}, \bibinfo {author} {\bibfnamefont {C.}~\bibnamefont {S\"underhauf}}, \bibinfo {author} {\bibfnamefont {N.}~\bibnamefont {Holzmann}}, \bibinfo {author} {\bibfnamefont {T.}~\bibnamefont {Ellaby}}, \bibinfo {author} {\bibfnamefont {R.~N.}\ \bibnamefont {Kerber}}, \bibinfo {author} {\bibfnamefont {G.}~\bibnamefont {Jones}},\ and\ \bibinfo {author} {\bibfnamefont {J.}~\bibnamefont {Camps}},\ }\bibfield  {title} {\bibinfo {title} {Quantum computation for periodic solids in second quantization},\ }\href {https://doi.org/10.1103/PhysRevResearch.5.013200} {\bibfield  {journal} {\bibinfo  {journal} {Phys. Rev. Res.}\ }\textbf {\bibinfo {volume} {5}},\ \bibinfo {pages} {013200} (\bibinfo {year} {2023})}\BibitemShut {NoStop}%
\bibitem [{\citenamefont {Ivanov}\ \emph {et~al.}(2024)\citenamefont {Ivanov}, \citenamefont {Patterson}, \citenamefont {Bothe}, \citenamefont {S{\"u}nderhauf}, \citenamefont {Berntson}, \citenamefont {Mortensen}, \citenamefont {Kuisma}, \citenamefont {Campbell},\ and\ \citenamefont {Izs{\'a}k}}]{ivanov2024quantum}%
  \BibitemOpen
  \bibfield  {author} {\bibinfo {author} {\bibfnamefont {A.~V.}\ \bibnamefont {Ivanov}}, \bibinfo {author} {\bibfnamefont {A.}~\bibnamefont {Patterson}}, \bibinfo {author} {\bibfnamefont {M.}~\bibnamefont {Bothe}}, \bibinfo {author} {\bibfnamefont {C.}~\bibnamefont {S{\"u}nderhauf}}, \bibinfo {author} {\bibfnamefont {B.~K.}\ \bibnamefont {Berntson}}, \bibinfo {author} {\bibfnamefont {J.~J.}\ \bibnamefont {Mortensen}}, \bibinfo {author} {\bibfnamefont {M.}~\bibnamefont {Kuisma}}, \bibinfo {author} {\bibfnamefont {E.}~\bibnamefont {Campbell}},\ and\ \bibinfo {author} {\bibfnamefont {R.}~\bibnamefont {Izs{\'a}k}},\ }\bibfield  {title} {\bibinfo {title} {Quantum computation of electronic structure with projector augmented-wave method and plane wave basis set},\ }\bibfield  {journal} {\bibinfo  {journal} {arXiv preprint arXiv:2408.03159}\ }\href {https://doi.org/10.48550/arXiv.2408.03159} {10.48550/arXiv.2408.03159} (\bibinfo {year} {2024})\BibitemShut {NoStop}%
\bibitem [{\citenamefont {Berry}\ \emph {et~al.}(2024)\citenamefont {Berry}, \citenamefont {Rubin}, \citenamefont {Elnabawy}, \citenamefont {Ahlers}, \citenamefont {DePrince~III}, \citenamefont {Lee}, \citenamefont {Gogolin},\ and\ \citenamefont {Babbush}}]{berry2024quantum}%
  \BibitemOpen
  \bibfield  {author} {\bibinfo {author} {\bibfnamefont {D.~W.}\ \bibnamefont {Berry}}, \bibinfo {author} {\bibfnamefont {N.~C.}\ \bibnamefont {Rubin}}, \bibinfo {author} {\bibfnamefont {A.~O.}\ \bibnamefont {Elnabawy}}, \bibinfo {author} {\bibfnamefont {G.}~\bibnamefont {Ahlers}}, \bibinfo {author} {\bibfnamefont {A.~E.}\ \bibnamefont {DePrince~III}}, \bibinfo {author} {\bibfnamefont {J.}~\bibnamefont {Lee}}, \bibinfo {author} {\bibfnamefont {C.}~\bibnamefont {Gogolin}},\ and\ \bibinfo {author} {\bibfnamefont {R.}~\bibnamefont {Babbush}},\ }\bibfield  {title} {\bibinfo {title} {Quantum simulation of realistic materials in first quantization using non-local pseudopotentials},\ }\href {https://doi.org/10.1038/s41534-024-00896-9} {\bibfield  {journal} {\bibinfo  {journal} {npj Quantum Information}\ }\textbf {\bibinfo {volume} {10}},\ \bibinfo {pages} {130} (\bibinfo {year} {2024})}\BibitemShut {NoStop}%
\bibitem [{\citenamefont {Georges}\ \emph {et~al.}(2024)\citenamefont {Georges}, \citenamefont {Bothe}, \citenamefont {S{\"u}nderhauf}, \citenamefont {Berntson}, \citenamefont {Izs{\'a}k},\ and\ \citenamefont {Ivanov}}]{georges2024quantum}%
  \BibitemOpen
  \bibfield  {author} {\bibinfo {author} {\bibfnamefont {T.~N.}\ \bibnamefont {Georges}}, \bibinfo {author} {\bibfnamefont {M.}~\bibnamefont {Bothe}}, \bibinfo {author} {\bibfnamefont {C.}~\bibnamefont {S{\"u}nderhauf}}, \bibinfo {author} {\bibfnamefont {B.~K.}\ \bibnamefont {Berntson}}, \bibinfo {author} {\bibfnamefont {R.}~\bibnamefont {Izs{\'a}k}},\ and\ \bibinfo {author} {\bibfnamefont {A.~V.}\ \bibnamefont {Ivanov}},\ }\bibfield  {title} {\bibinfo {title} {Quantum simulations of chemistry in first quantization with any basis set},\ }\bibfield  {journal} {\bibinfo  {journal} {arXiv preprint arXiv:2408.03145}\ }\href {https://doi.org/10.48550/arXiv.2408.03145} {10.48550/arXiv.2408.03145} (\bibinfo {year} {2024})\BibitemShut {NoStop}%
\bibitem [{\citenamefont {McClean}\ \emph {et~al.}(2016)\citenamefont {McClean}, \citenamefont {Romero}, \citenamefont {Babbush},\ and\ \citenamefont {Aspuru-Guzik}}]{vqe_theory}%
  \BibitemOpen
  \bibfield  {author} {\bibinfo {author} {\bibfnamefont {J.~R.}\ \bibnamefont {McClean}}, \bibinfo {author} {\bibfnamefont {J.}~\bibnamefont {Romero}}, \bibinfo {author} {\bibfnamefont {R.}~\bibnamefont {Babbush}},\ and\ \bibinfo {author} {\bibfnamefont {A.}~\bibnamefont {Aspuru-Guzik}},\ }\bibfield  {title} {\bibinfo {title} {The theory of variational hybrid quantum-classical algorithms},\ }\href {https://doi.org/10.1088/1367-2630/18/2/023023} {\bibfield  {journal} {\bibinfo  {journal} {New J. Phys.}\ }\textbf {\bibinfo {volume} {18}},\ \bibinfo {pages} {023023} (\bibinfo {year} {2016})}\BibitemShut {NoStop}%
\bibitem [{\citenamefont {Clinton}\ \emph {et~al.}(2024)\citenamefont {Clinton}, \citenamefont {Cubitt}, \citenamefont {Flynn}, \citenamefont {Gambetta}, \citenamefont {Klassen}, \citenamefont {Montanaro}, \citenamefont {Piddock}, \citenamefont {Santos},\ and\ \citenamefont {Sheridan}}]{clinton2024towards}%
  \BibitemOpen
  \bibfield  {author} {\bibinfo {author} {\bibfnamefont {L.}~\bibnamefont {Clinton}}, \bibinfo {author} {\bibfnamefont {T.}~\bibnamefont {Cubitt}}, \bibinfo {author} {\bibfnamefont {B.}~\bibnamefont {Flynn}}, \bibinfo {author} {\bibfnamefont {F.~M.}\ \bibnamefont {Gambetta}}, \bibinfo {author} {\bibfnamefont {J.}~\bibnamefont {Klassen}}, \bibinfo {author} {\bibfnamefont {A.}~\bibnamefont {Montanaro}}, \bibinfo {author} {\bibfnamefont {S.}~\bibnamefont {Piddock}}, \bibinfo {author} {\bibfnamefont {R.~A.}\ \bibnamefont {Santos}},\ and\ \bibinfo {author} {\bibfnamefont {E.}~\bibnamefont {Sheridan}},\ }\bibfield  {title} {\bibinfo {title} {Towards near-term quantum simulation of materials},\ }\href {https://doi.org/10.1038/s41467-023-43479-6} {\bibfield  {journal} {\bibinfo  {journal} {Nature Communications}\ }\textbf {\bibinfo {volume} {15}},\ \bibinfo {pages} {211} (\bibinfo {year} {2024})}\BibitemShut {NoStop}%
\bibitem [{\citenamefont {Georges}\ \emph {et~al.}(1996)\citenamefont {Georges}, \citenamefont {Kotliar}, \citenamefont {Krauth},\ and\ \citenamefont {Rozenberg}}]{dmft_georges96}%
  \BibitemOpen
  \bibfield  {author} {\bibinfo {author} {\bibfnamefont {A.}~\bibnamefont {Georges}}, \bibinfo {author} {\bibfnamefont {G.}~\bibnamefont {Kotliar}}, \bibinfo {author} {\bibfnamefont {W.}~\bibnamefont {Krauth}},\ and\ \bibinfo {author} {\bibfnamefont {M.~J.}\ \bibnamefont {Rozenberg}},\ }\bibfield  {title} {\bibinfo {title} {Dynamical mean-field theory of strongly correlated fermion systems and the limit of infinite dimensions},\ }\href {https://doi.org/10.1103/RevModPhys.68.13} {\bibfield  {journal} {\bibinfo  {journal} {Rev. Mod. Phys.}\ }\textbf {\bibinfo {volume} {68}},\ \bibinfo {pages} {13} (\bibinfo {year} {1996})}\BibitemShut {NoStop}%
\bibitem [{\citenamefont {Held}(2007)}]{dmft_held07}%
  \BibitemOpen
  \bibfield  {author} {\bibinfo {author} {\bibfnamefont {K.}~\bibnamefont {Held}},\ }\bibfield  {title} {\bibinfo {title} {Electronic structure calculations using dynamical mean field theory},\ }\href {https://doi.org/10.1080/00018730701619647} {\bibfield  {journal} {\bibinfo  {journal} {Adv. Phys.}\ }\textbf {\bibinfo {volume} {56}},\ \bibinfo {pages} {829} (\bibinfo {year} {2007})}\BibitemShut {NoStop}%
\bibitem [{\citenamefont {Bauer}\ \emph {et~al.}(2016)\citenamefont {Bauer}, \citenamefont {Wecker}, \citenamefont {Millis}, \citenamefont {Hastings},\ and\ \citenamefont {Troyer}}]{hybrd_dmft}%
  \BibitemOpen
  \bibfield  {author} {\bibinfo {author} {\bibfnamefont {B.}~\bibnamefont {Bauer}}, \bibinfo {author} {\bibfnamefont {D.}~\bibnamefont {Wecker}}, \bibinfo {author} {\bibfnamefont {A.~J.}\ \bibnamefont {Millis}}, \bibinfo {author} {\bibfnamefont {M.~B.}\ \bibnamefont {Hastings}},\ and\ \bibinfo {author} {\bibfnamefont {M.}~\bibnamefont {Troyer}},\ }\bibfield  {title} {\bibinfo {title} {Hybrid quantum-classical approach to correlated materials},\ }\href {https://doi.org/10.1103/PhysRevX.6.031045} {\bibfield  {journal} {\bibinfo  {journal} {Phys. Rev. X}\ }\textbf {\bibinfo {volume} {6}},\ \bibinfo {pages} {031045} (\bibinfo {year} {2016})}\BibitemShut {NoStop}%
\bibitem [{\citenamefont {Rungger}\ \emph {et~al.}(2019)\citenamefont {Rungger}, \citenamefont {Fitzpatrick}, \citenamefont {Chen}, \citenamefont {Alderete}, \citenamefont {Apel}, \citenamefont {Cowtan}, \citenamefont {Patterson}, \citenamefont {Ramo}, \citenamefont {Zhu}, \citenamefont {Nguyen} \emph {et~al.}}]{hybrd_2sitedmft2}%
  \BibitemOpen
  \bibfield  {author} {\bibinfo {author} {\bibfnamefont {I.}~\bibnamefont {Rungger}}, \bibinfo {author} {\bibfnamefont {N.}~\bibnamefont {Fitzpatrick}}, \bibinfo {author} {\bibfnamefont {H.}~\bibnamefont {Chen}}, \bibinfo {author} {\bibfnamefont {C.}~\bibnamefont {Alderete}}, \bibinfo {author} {\bibfnamefont {H.}~\bibnamefont {Apel}}, \bibinfo {author} {\bibfnamefont {A.}~\bibnamefont {Cowtan}}, \bibinfo {author} {\bibfnamefont {A.}~\bibnamefont {Patterson}}, \bibinfo {author} {\bibfnamefont {D.~M.}\ \bibnamefont {Ramo}}, \bibinfo {author} {\bibfnamefont {Y.}~\bibnamefont {Zhu}}, \bibinfo {author} {\bibfnamefont {N.}~\bibnamefont {Nguyen}}, \emph {et~al.},\ }\bibfield  {title} {\bibinfo {title} {Dynamical mean field theory algorithm and experiment on quantum computers},\ }\bibfield  {journal} {\bibinfo  {journal} {arXiv:1910.04735}\ }\href {https://doi.org/10.48550/arXiv.1910.04735} {10.48550/arXiv.1910.04735} (\bibinfo {year} {2019})\BibitemShut {NoStop}%
\bibitem [{\citenamefont {Backes}\ \emph {et~al.}(2023)\citenamefont {Backes}, \citenamefont {Murakami}, \citenamefont {Sakai},\ and\ \citenamefont {Arita}}]{Backes2023DMFT}%
  \BibitemOpen
  \bibfield  {author} {\bibinfo {author} {\bibfnamefont {S.}~\bibnamefont {Backes}}, \bibinfo {author} {\bibfnamefont {Y.}~\bibnamefont {Murakami}}, \bibinfo {author} {\bibfnamefont {S.}~\bibnamefont {Sakai}},\ and\ \bibinfo {author} {\bibfnamefont {R.}~\bibnamefont {Arita}},\ }\bibfield  {title} {\bibinfo {title} {Dynamical mean-field theory for the hubbard-holstein model on a quantum device},\ }\href {https://doi.org/10.1103/PhysRevB.107.165155} {\bibfield  {journal} {\bibinfo  {journal} {Phys. Rev. B}\ }\textbf {\bibinfo {volume} {107}},\ \bibinfo {pages} {165155} (\bibinfo {year} {2023})}\BibitemShut {NoStop}%
\bibitem [{\citenamefont {Selisko}\ \emph {et~al.}(2024)\citenamefont {Selisko}, \citenamefont {Amsler}, \citenamefont {Wever}, \citenamefont {Kawashima}, \citenamefont {Samsonidze}, \citenamefont {Haq}, \citenamefont {Tacchino}, \citenamefont {Tavernelli},\ and\ \citenamefont {Eckl}}]{selisko2024DMFT}%
  \BibitemOpen
  \bibfield  {author} {\bibinfo {author} {\bibfnamefont {J.}~\bibnamefont {Selisko}}, \bibinfo {author} {\bibfnamefont {M.}~\bibnamefont {Amsler}}, \bibinfo {author} {\bibfnamefont {C.}~\bibnamefont {Wever}}, \bibinfo {author} {\bibfnamefont {Y.}~\bibnamefont {Kawashima}}, \bibinfo {author} {\bibfnamefont {G.}~\bibnamefont {Samsonidze}}, \bibinfo {author} {\bibfnamefont {R.~U.}\ \bibnamefont {Haq}}, \bibinfo {author} {\bibfnamefont {F.}~\bibnamefont {Tacchino}}, \bibinfo {author} {\bibfnamefont {I.}~\bibnamefont {Tavernelli}},\ and\ \bibinfo {author} {\bibfnamefont {T.}~\bibnamefont {Eckl}},\ }\bibfield  {title} {\bibinfo {title} {Dynamical mean field theory for real materials on a quantum computer},\ }\bibfield  {journal} {\bibinfo  {journal} {arXiv preprint arXiv:2404.09527}\ }\href {https://doi.org/10.48550/arXiv.2404.09527} {10.48550/arXiv.2404.09527} (\bibinfo {year} {2024})\BibitemShut {NoStop}%
\bibitem [{\citenamefont {Lanata}\ \emph {et~al.}(2013)\citenamefont {Lanata}, \citenamefont {Yao}, \citenamefont {Wang}, \citenamefont {Ho}, \citenamefont {Schmalian}, \citenamefont {Haule},\ and\ \citenamefont {Kotliar}}]{ga_ce}%
  \BibitemOpen
  \bibfield  {author} {\bibinfo {author} {\bibfnamefont {N.}~\bibnamefont {Lanata}}, \bibinfo {author} {\bibfnamefont {Y.-X.}\ \bibnamefont {Yao}}, \bibinfo {author} {\bibfnamefont {C.-Z.}\ \bibnamefont {Wang}}, \bibinfo {author} {\bibfnamefont {K.-M.}\ \bibnamefont {Ho}}, \bibinfo {author} {\bibfnamefont {J.}~\bibnamefont {Schmalian}}, \bibinfo {author} {\bibfnamefont {K.}~\bibnamefont {Haule}},\ and\ \bibinfo {author} {\bibfnamefont {G.}~\bibnamefont {Kotliar}},\ }\bibfield  {title} {\bibinfo {title} {$\gamma$- $\alpha$ isostructural transition in cerium},\ }\href {https://doi.org/10.1103/PhysRevLett.111.196801} {\bibfield  {journal} {\bibinfo  {journal} {Phys. Rev. Lett.}\ }\textbf {\bibinfo {volume} {111}},\ \bibinfo {pages} {196801} (\bibinfo {year} {2013})}\BibitemShut {NoStop}%
\bibitem [{\citenamefont {Deng}\ \emph {et~al.}(2008)\citenamefont {Deng}, \citenamefont {Dai},\ and\ \citenamefont {Fang}}]{deng2008lda+}%
  \BibitemOpen
  \bibfield  {author} {\bibinfo {author} {\bibfnamefont {X.}~\bibnamefont {Deng}}, \bibinfo {author} {\bibfnamefont {X.}~\bibnamefont {Dai}},\ and\ \bibinfo {author} {\bibfnamefont {Z.}~\bibnamefont {Fang}},\ }\bibfield  {title} {\bibinfo {title} {{LDA + Gutzwiller method for correlated electron systems}},\ }\href {https://doi.org/10.1209/0295-5075/83/37008} {\bibfield  {journal} {\bibinfo  {journal} {EPL}\ }\textbf {\bibinfo {volume} {83}},\ \bibinfo {pages} {37008} (\bibinfo {year} {2008})}\BibitemShut {NoStop}%
\bibitem [{\citenamefont {Lanat{\`a}}\ \emph {et~al.}(2015)\citenamefont {Lanat{\`a}}, \citenamefont {Yao}, \citenamefont {Wang}, \citenamefont {Ho},\ and\ \citenamefont {Kotliar}}]{ga_pu}%
  \BibitemOpen
  \bibfield  {author} {\bibinfo {author} {\bibfnamefont {N.}~\bibnamefont {Lanat{\`a}}}, \bibinfo {author} {\bibfnamefont {Y.-X.}\ \bibnamefont {Yao}}, \bibinfo {author} {\bibfnamefont {C.-Z.}\ \bibnamefont {Wang}}, \bibinfo {author} {\bibfnamefont {K.-M.}\ \bibnamefont {Ho}},\ and\ \bibinfo {author} {\bibfnamefont {G.}~\bibnamefont {Kotliar}},\ }\bibfield  {title} {\bibinfo {title} {Phase diagram and electronic structure of praseodymium and plutonium},\ }\href {https://doi.org/10.1103/PhysRevX.5.011008} {\bibfield  {journal} {\bibinfo  {journal} {Phys. Rev. X}\ }\textbf {\bibinfo {volume} {5}},\ \bibinfo {pages} {011008} (\bibinfo {year} {2015})}\BibitemShut {NoStop}%
\bibitem [{\citenamefont {Lanat\`a}\ \emph {et~al.}(2017)\citenamefont {Lanat\`a}, \citenamefont {Yao}, \citenamefont {Deng}, \citenamefont {Dobrosavljevi\ifmmode~\acute{c}\else \'{c}\fi{}},\ and\ \citenamefont {Kotliar}}]{ga_uo2}%
  \BibitemOpen
  \bibfield  {author} {\bibinfo {author} {\bibfnamefont {N.}~\bibnamefont {Lanat\`a}}, \bibinfo {author} {\bibfnamefont {Y.-X.}\ \bibnamefont {Yao}}, \bibinfo {author} {\bibfnamefont {X.}~\bibnamefont {Deng}}, \bibinfo {author} {\bibfnamefont {V.}~\bibnamefont {Dobrosavljevi\ifmmode~\acute{c}\else \'{c}\fi{}}},\ and\ \bibinfo {author} {\bibfnamefont {G.}~\bibnamefont {Kotliar}},\ }\bibfield  {title} {\bibinfo {title} {Slave boson theory of orbital differentiation with crystal field effects: Application to {{${\mathrm{UO}}_{2}$}}},\ }\href {https://doi.org/10.1103/PhysRevLett.118.126401} {\bibfield  {journal} {\bibinfo  {journal} {Phys. Rev. Lett.}\ }\textbf {\bibinfo {volume} {118}},\ \bibinfo {pages} {126401} (\bibinfo {year} {2017})}\BibitemShut {NoStop}%
\bibitem [{\citenamefont {Lanat{\`a}}\ \emph {et~al.}(2019)\citenamefont {Lanat{\`a}}, \citenamefont {Lee}, \citenamefont {Yao}, \citenamefont {Stevanovi{\'c}},\ and\ \citenamefont {Dobrosavljevi{\'c}}}]{ga_tmo}%
  \BibitemOpen
  \bibfield  {author} {\bibinfo {author} {\bibfnamefont {N.}~\bibnamefont {Lanat{\`a}}}, \bibinfo {author} {\bibfnamefont {T.-H.}\ \bibnamefont {Lee}}, \bibinfo {author} {\bibfnamefont {Y.-X.}\ \bibnamefont {Yao}}, \bibinfo {author} {\bibfnamefont {V.}~\bibnamefont {Stevanovi{\'c}}},\ and\ \bibinfo {author} {\bibfnamefont {V.}~\bibnamefont {Dobrosavljevi{\'c}}},\ }\bibfield  {title} {\bibinfo {title} {Connection between mott physics and crystal structure in a series of transition metal binary compounds},\ }\href {https://doi.org/10.1038/s41524-019-0169-0} {\bibfield  {journal} {\bibinfo  {journal} {npj Comput. Mater.}\ }\textbf {\bibinfo {volume} {5}},\ \bibinfo {pages} {30} (\bibinfo {year} {2019})}\BibitemShut {NoStop}%
\bibitem [{\citenamefont {Yao}\ \emph {et~al.}(2021{\natexlab{a}})\citenamefont {Yao}, \citenamefont {Zhang}, \citenamefont {Wang}, \citenamefont {Ho},\ and\ \citenamefont {Orth}}]{gqce}%
  \BibitemOpen
  \bibfield  {author} {\bibinfo {author} {\bibfnamefont {Y.-X.}\ \bibnamefont {Yao}}, \bibinfo {author} {\bibfnamefont {F.}~\bibnamefont {Zhang}}, \bibinfo {author} {\bibfnamefont {C.-Z.}\ \bibnamefont {Wang}}, \bibinfo {author} {\bibfnamefont {K.-M.}\ \bibnamefont {Ho}},\ and\ \bibinfo {author} {\bibfnamefont {P.~P.}\ \bibnamefont {Orth}},\ }\bibfield  {title} {\bibinfo {title} {Gutzwiller hybrid quantum-classical computing approach for correlated materials},\ }\href {https://doi.org/10.1103/PhysRevResearch.3.013184} {\bibfield  {journal} {\bibinfo  {journal} {Phys. Rev. Res.}\ }\textbf {\bibinfo {volume} {3}},\ \bibinfo {pages} {013184} (\bibinfo {year} {2021}{\natexlab{a}})}\BibitemShut {NoStop}%
\bibitem [{\citenamefont {Mukherjee}\ \emph {et~al.}(2023)\citenamefont {Mukherjee}, \citenamefont {Berthusen}, \citenamefont {Getelina}, \citenamefont {Orth},\ and\ \citenamefont {Yao}}]{mukherjee2023comparative}%
  \BibitemOpen
  \bibfield  {author} {\bibinfo {author} {\bibfnamefont {A.}~\bibnamefont {Mukherjee}}, \bibinfo {author} {\bibfnamefont {N.~F.}\ \bibnamefont {Berthusen}}, \bibinfo {author} {\bibfnamefont {J.~C.}\ \bibnamefont {Getelina}}, \bibinfo {author} {\bibfnamefont {P.~P.}\ \bibnamefont {Orth}},\ and\ \bibinfo {author} {\bibfnamefont {Y.-X.}\ \bibnamefont {Yao}},\ }\bibfield  {title} {\bibinfo {title} {Comparative study of adaptive variational quantum eigensolvers for multi-orbital impurity models},\ }\href {https://doi.org/10.1038/s42005-022-01089-6} {\bibfield  {journal} {\bibinfo  {journal} {Commun. Phys.}\ }\textbf {\bibinfo {volume} {6}},\ \bibinfo {pages} {4} (\bibinfo {year} {2023})}\BibitemShut {NoStop}%
\bibitem [{\citenamefont {Lee}\ \emph {et~al.}(2019)\citenamefont {Lee}, \citenamefont {Ayral}, \citenamefont {Yao}, \citenamefont {Lanata},\ and\ \citenamefont {Kotliar}}]{ga_dmet}%
  \BibitemOpen
  \bibfield  {author} {\bibinfo {author} {\bibfnamefont {T.-H.}\ \bibnamefont {Lee}}, \bibinfo {author} {\bibfnamefont {T.}~\bibnamefont {Ayral}}, \bibinfo {author} {\bibfnamefont {Y.-X.}\ \bibnamefont {Yao}}, \bibinfo {author} {\bibfnamefont {N.}~\bibnamefont {Lanata}},\ and\ \bibinfo {author} {\bibfnamefont {G.}~\bibnamefont {Kotliar}},\ }\bibfield  {title} {\bibinfo {title} {Rotationally invariant slave-boson and density matrix embedding theory: Unified framework and comparative study on the one-dimensional and two-dimensional hubbard model},\ }\href {https://doi.org/10.1103/PhysRevB.99.115129} {\bibfield  {journal} {\bibinfo  {journal} {Phys. Rev. B}\ }\textbf {\bibinfo {volume} {99}},\ \bibinfo {pages} {115129} (\bibinfo {year} {2019})}\BibitemShut {NoStop}%
\bibitem [{\citenamefont {Mineh}\ and\ \citenamefont {Montanaro}(2022)}]{Mineh2022DMET}%
  \BibitemOpen
  \bibfield  {author} {\bibinfo {author} {\bibfnamefont {L.}~\bibnamefont {Mineh}}\ and\ \bibinfo {author} {\bibfnamefont {A.}~\bibnamefont {Montanaro}},\ }\bibfield  {title} {\bibinfo {title} {Solving the hubbard model using density matrix embedding theory and the variational quantum eigensolver},\ }\href {https://doi.org/10.1103/PhysRevB.105.125117} {\bibfield  {journal} {\bibinfo  {journal} {Phys. Rev. B}\ }\textbf {\bibinfo {volume} {105}},\ \bibinfo {pages} {125117} (\bibinfo {year} {2022})}\BibitemShut {NoStop}%
\bibitem [{\citenamefont {Kawashima}\ \emph {et~al.}(2021)\citenamefont {Kawashima}, \citenamefont {Lloyd}, \citenamefont {Coons}, \citenamefont {Nam}, \citenamefont {Matsuura}, \citenamefont {Garza}, \citenamefont {Johri}, \citenamefont {Huntington}, \citenamefont {Senicourt}, \citenamefont {Maksymov} \emph {et~al.}}]{kawashima2021DMET}%
  \BibitemOpen
  \bibfield  {author} {\bibinfo {author} {\bibfnamefont {Y.}~\bibnamefont {Kawashima}}, \bibinfo {author} {\bibfnamefont {E.}~\bibnamefont {Lloyd}}, \bibinfo {author} {\bibfnamefont {M.~P.}\ \bibnamefont {Coons}}, \bibinfo {author} {\bibfnamefont {Y.}~\bibnamefont {Nam}}, \bibinfo {author} {\bibfnamefont {S.}~\bibnamefont {Matsuura}}, \bibinfo {author} {\bibfnamefont {A.~J.}\ \bibnamefont {Garza}}, \bibinfo {author} {\bibfnamefont {S.}~\bibnamefont {Johri}}, \bibinfo {author} {\bibfnamefont {L.}~\bibnamefont {Huntington}}, \bibinfo {author} {\bibfnamefont {V.}~\bibnamefont {Senicourt}}, \bibinfo {author} {\bibfnamefont {A.~O.}\ \bibnamefont {Maksymov}}, \emph {et~al.},\ }\bibfield  {title} {\bibinfo {title} {Optimizing electronic structure simulations on a trapped-ion quantum computer using problem decomposition},\ }\href {https://doi.org/10.1038/s42005-021-00751-9} {\bibfield  {journal} {\bibinfo  {journal} {Communications Physics}\ }\textbf {\bibinfo {volume} {4}},\ \bibinfo {pages} {245} (\bibinfo {year}
  {2021})}\BibitemShut {NoStop}%
\bibitem [{\citenamefont {Rubin}(2016)}]{rubin2016DMET}%
  \BibitemOpen
  \bibfield  {author} {\bibinfo {author} {\bibfnamefont {N.~C.}\ \bibnamefont {Rubin}},\ }\bibfield  {title} {\bibinfo {title} {A hybrid classical/quantum approach for large-scale studies of quantum systems with density matrix embedding theory},\ }\bibfield  {journal} {\bibinfo  {journal} {arXiv preprint arXiv:1610.06910}\ }\href {https://doi.org/10.48550/arXiv.1610.06910} {10.48550/arXiv.1610.06910} (\bibinfo {year} {2016})\BibitemShut {NoStop}%
\bibitem [{\citenamefont {Lanata}\ \emph {et~al.}(2017)\citenamefont {Lanata}, \citenamefont {Lee}, \citenamefont {Yao},\ and\ \citenamefont {Dobrosavljevi{\'c}}}]{gga_lanata2017}%
  \BibitemOpen
  \bibfield  {author} {\bibinfo {author} {\bibfnamefont {N.}~\bibnamefont {Lanata}}, \bibinfo {author} {\bibfnamefont {T.-H.}\ \bibnamefont {Lee}}, \bibinfo {author} {\bibfnamefont {Y.-X.}\ \bibnamefont {Yao}},\ and\ \bibinfo {author} {\bibfnamefont {V.}~\bibnamefont {Dobrosavljevi{\'c}}},\ }\bibfield  {title} {\bibinfo {title} {Emergent bloch excitations in mott matter},\ }\href@noop {} {\bibfield  {journal} {\bibinfo  {journal} {Phys. Rev. B}\ }\textbf {\bibinfo {volume} {96}},\ \bibinfo {pages} {195126} (\bibinfo {year} {2017})}\BibitemShut {NoStop}%
\bibitem [{\citenamefont {Frank}\ \emph {et~al.}(2021)\citenamefont {Frank}, \citenamefont {Lee}, \citenamefont {Bhattacharyya}, \citenamefont {Tsang}, \citenamefont {Quito}, \citenamefont {Dobrosavljevi\ifmmode~\acute{c}\else \'{c}\fi{}}, \citenamefont {Christiansen},\ and\ \citenamefont {Lanat\`a}}]{Frank2021QuantumED}%
  \BibitemOpen
  \bibfield  {author} {\bibinfo {author} {\bibfnamefont {M.~S.}\ \bibnamefont {Frank}}, \bibinfo {author} {\bibfnamefont {T.-H.}\ \bibnamefont {Lee}}, \bibinfo {author} {\bibfnamefont {G.}~\bibnamefont {Bhattacharyya}}, \bibinfo {author} {\bibfnamefont {P.~K.~H.}\ \bibnamefont {Tsang}}, \bibinfo {author} {\bibfnamefont {V.~L.}\ \bibnamefont {Quito}}, \bibinfo {author} {\bibfnamefont {V.}~\bibnamefont {Dobrosavljevi\ifmmode~\acute{c}\else \'{c}\fi{}}}, \bibinfo {author} {\bibfnamefont {O.}~\bibnamefont {Christiansen}},\ and\ \bibinfo {author} {\bibfnamefont {N.}~\bibnamefont {Lanat\`a}},\ }\bibfield  {title} {\bibinfo {title} {Quantum embedding description of the anderson lattice model with the ghost gutzwiller approximation},\ }\href {https://doi.org/10.1103/PhysRevB.104.L081103} {\bibfield  {journal} {\bibinfo  {journal} {Phys. Rev. B}\ }\textbf {\bibinfo {volume} {104}},\ \bibinfo {pages} {L081103} (\bibinfo {year} {2021})}\BibitemShut {NoStop}%
\bibitem [{\citenamefont {Lee}\ \emph {et~al.}(2024)\citenamefont {Lee}, \citenamefont {Melnick}, \citenamefont {Adler}, \citenamefont {Sun}, \citenamefont {Yao}, \citenamefont {Lanat\`a},\ and\ \citenamefont {Kotliar}}]{Lee2024ChargeSC}%
  \BibitemOpen
  \bibfield  {author} {\bibinfo {author} {\bibfnamefont {T.-H.}\ \bibnamefont {Lee}}, \bibinfo {author} {\bibfnamefont {C.}~\bibnamefont {Melnick}}, \bibinfo {author} {\bibfnamefont {R.}~\bibnamefont {Adler}}, \bibinfo {author} {\bibfnamefont {X.}~\bibnamefont {Sun}}, \bibinfo {author} {\bibfnamefont {Y.-X.}\ \bibnamefont {Yao}}, \bibinfo {author} {\bibfnamefont {N.}~\bibnamefont {Lanat\`a}},\ and\ \bibinfo {author} {\bibfnamefont {G.}~\bibnamefont {Kotliar}},\ }\bibfield  {title} {\bibinfo {title} {Charge self-consistent density functional theory plus ghost rotationally invariant slave-boson theory for correlated materials},\ }\href {https://doi.org/10.1103/PhysRevB.110.115126} {\bibfield  {journal} {\bibinfo  {journal} {Phys. Rev. B}\ }\textbf {\bibinfo {volume} {110}},\ \bibinfo {pages} {115126} (\bibinfo {year} {2024})}\BibitemShut {NoStop}%
\bibitem [{\citenamefont {Gomes}\ \emph {et~al.}(2021)\citenamefont {Gomes}, \citenamefont {Mukherjee}, \citenamefont {Zhang}, \citenamefont {Iadecola}, \citenamefont {Wang}, \citenamefont {Ho}, \citenamefont {Orth},\ and\ \citenamefont {Yao}}]{AVQITE}%
  \BibitemOpen
  \bibfield  {author} {\bibinfo {author} {\bibfnamefont {N.}~\bibnamefont {Gomes}}, \bibinfo {author} {\bibfnamefont {A.}~\bibnamefont {Mukherjee}}, \bibinfo {author} {\bibfnamefont {F.}~\bibnamefont {Zhang}}, \bibinfo {author} {\bibfnamefont {T.}~\bibnamefont {Iadecola}}, \bibinfo {author} {\bibfnamefont {C.-Z.}\ \bibnamefont {Wang}}, \bibinfo {author} {\bibfnamefont {K.-M.}\ \bibnamefont {Ho}}, \bibinfo {author} {\bibfnamefont {P.~P.}\ \bibnamefont {Orth}},\ and\ \bibinfo {author} {\bibfnamefont {Y.-X.}\ \bibnamefont {Yao}},\ }\bibfield  {title} {\bibinfo {title} {Adaptive variational quantum imaginary time evolution approach for ground state preparation},\ }\href {https://doi.org/10.1002/qute.202100114} {\bibfield  {journal} {\bibinfo  {journal} {Adv. Quantum Technol.}\ }\textbf {\bibinfo {volume} {4}},\ \bibinfo {pages} {2100114} (\bibinfo {year} {2021})}\BibitemShut {NoStop}%
\bibitem [{\citenamefont {Yao}\ \emph {et~al.}(2021{\natexlab{b}})\citenamefont {Yao}, \citenamefont {Gomes}, \citenamefont {Zhang}, \citenamefont {Wang}, \citenamefont {Ho}, \citenamefont {Iadecola},\ and\ \citenamefont {Orth}}]{AVQDS}%
  \BibitemOpen
  \bibfield  {author} {\bibinfo {author} {\bibfnamefont {Y.-X.}\ \bibnamefont {Yao}}, \bibinfo {author} {\bibfnamefont {N.}~\bibnamefont {Gomes}}, \bibinfo {author} {\bibfnamefont {F.}~\bibnamefont {Zhang}}, \bibinfo {author} {\bibfnamefont {C.-Z.}\ \bibnamefont {Wang}}, \bibinfo {author} {\bibfnamefont {K.-M.}\ \bibnamefont {Ho}}, \bibinfo {author} {\bibfnamefont {T.}~\bibnamefont {Iadecola}},\ and\ \bibinfo {author} {\bibfnamefont {P.~P.}\ \bibnamefont {Orth}},\ }\bibfield  {title} {\bibinfo {title} {Adaptive variational quantum dynamics simulations},\ }\href {https://doi.org/10.1103/PRXQuantum.2.030307} {\bibfield  {journal} {\bibinfo  {journal} {PRX Quantum}\ }\textbf {\bibinfo {volume} {2}},\ \bibinfo {pages} {030307} (\bibinfo {year} {2021}{\natexlab{b}})}\BibitemShut {NoStop}%
\bibitem [{\citenamefont {Frank}\ \emph {et~al.}(2024)\citenamefont {Frank}, \citenamefont {Artiukhin}, \citenamefont {Lee}, \citenamefont {Yao}, \citenamefont {Barros}, \citenamefont {Christiansen},\ and\ \citenamefont {Lanat{\`a}}}]{Frank2024activeLA}%
  \BibitemOpen
  \bibfield  {author} {\bibinfo {author} {\bibfnamefont {M.~S.}\ \bibnamefont {Frank}}, \bibinfo {author} {\bibfnamefont {D.~G.}\ \bibnamefont {Artiukhin}}, \bibinfo {author} {\bibfnamefont {T.-H.}\ \bibnamefont {Lee}}, \bibinfo {author} {\bibfnamefont {Y.-X.}\ \bibnamefont {Yao}}, \bibinfo {author} {\bibfnamefont {K.}~\bibnamefont {Barros}}, \bibinfo {author} {\bibfnamefont {O.}~\bibnamefont {Christiansen}},\ and\ \bibinfo {author} {\bibfnamefont {N.}~\bibnamefont {Lanat{\`a}}},\ }\bibfield  {title} {\bibinfo {title} {Active learning approach to simulations of strongly correlated matter with the ghost gutzwiller approximation},\ }\href {https://doi.org/10.1103/PhysRevResearch.6.013242} {\bibfield  {journal} {\bibinfo  {journal} {Physical Review Research}\ }\textbf {\bibinfo {volume} {6}},\ \bibinfo {pages} {013242} (\bibinfo {year} {2024})}\BibitemShut {NoStop}%
\bibitem [{\citenamefont {Lee}\ \emph {et~al.}(2023)\citenamefont {Lee}, \citenamefont {Lanat\`a},\ and\ \citenamefont {Kotliar}}]{Lee2023AccuracyGRISB}%
  \BibitemOpen
  \bibfield  {author} {\bibinfo {author} {\bibfnamefont {T.-H.}\ \bibnamefont {Lee}}, \bibinfo {author} {\bibfnamefont {N.}~\bibnamefont {Lanat\`a}},\ and\ \bibinfo {author} {\bibfnamefont {G.}~\bibnamefont {Kotliar}},\ }\bibfield  {title} {\bibinfo {title} {Accuracy of ghost rotationally invariant slave-boson and dynamical mean field theory as a function of the impurity-model bath size},\ }\href {https://doi.org/10.1103/PhysRevB.107.L121104} {\bibfield  {journal} {\bibinfo  {journal} {Phys. Rev. B}\ }\textbf {\bibinfo {volume} {107}},\ \bibinfo {pages} {L121104} (\bibinfo {year} {2023})}\BibitemShut {NoStop}%
\bibitem [{\citenamefont {Savrasov}\ \emph {et~al.}(2006)\citenamefont {Savrasov}, \citenamefont {Haule},\ and\ \citenamefont {Kotliar}}]{Savrasov2006ManyBodyES}%
  \BibitemOpen
  \bibfield  {author} {\bibinfo {author} {\bibfnamefont {S.~Y.}\ \bibnamefont {Savrasov}}, \bibinfo {author} {\bibfnamefont {K.}~\bibnamefont {Haule}},\ and\ \bibinfo {author} {\bibfnamefont {G.}~\bibnamefont {Kotliar}},\ }\bibfield  {title} {\bibinfo {title} {Many-body electronic structure of americium metal},\ }\href {https://doi.org/10.1103/PhysRevLett.96.036404} {\bibfield  {journal} {\bibinfo  {journal} {Phys. Rev. Lett.}\ }\textbf {\bibinfo {volume} {96}},\ \bibinfo {pages} {036404} (\bibinfo {year} {2006})}\BibitemShut {NoStop}%
\bibitem [{\citenamefont {Bradbury}\ \emph {et~al.}(2018)\citenamefont {Bradbury}, \citenamefont {Frostig}, \citenamefont {Hawkins}, \citenamefont {Johnson}, \citenamefont {Leary}, \citenamefont {Maclaurin}, \citenamefont {Necula}, \citenamefont {Paszke}, \citenamefont {Vander{P}las}, \citenamefont {Wanderman-{M}ilne},\ and\ \citenamefont {Zhang}}]{jax2018github}%
  \BibitemOpen
  \bibfield  {author} {\bibinfo {author} {\bibfnamefont {J.}~\bibnamefont {Bradbury}}, \bibinfo {author} {\bibfnamefont {R.}~\bibnamefont {Frostig}}, \bibinfo {author} {\bibfnamefont {P.}~\bibnamefont {Hawkins}}, \bibinfo {author} {\bibfnamefont {M.~J.}\ \bibnamefont {Johnson}}, \bibinfo {author} {\bibfnamefont {C.}~\bibnamefont {Leary}}, \bibinfo {author} {\bibfnamefont {D.}~\bibnamefont {Maclaurin}}, \bibinfo {author} {\bibfnamefont {G.}~\bibnamefont {Necula}}, \bibinfo {author} {\bibfnamefont {A.}~\bibnamefont {Paszke}}, \bibinfo {author} {\bibfnamefont {J.}~\bibnamefont {Vander{P}las}}, \bibinfo {author} {\bibfnamefont {S.}~\bibnamefont {Wanderman-{M}ilne}},\ and\ \bibinfo {author} {\bibfnamefont {Q.}~\bibnamefont {Zhang}},\ }\href {http://github.com/jax-ml/jax} {\bibinfo {title} {{JAX}: composable transformations of {P}ython+{N}um{P}y programs}} (\bibinfo {year} {2018})\BibitemShut {NoStop}%
\bibitem [{\citenamefont {McArdle}\ \emph {et~al.}(2019)\citenamefont {McArdle}, \citenamefont {Jones}, \citenamefont {Endo}, \citenamefont {Li}, \citenamefont {Benjamin},\ and\ \citenamefont {Yuan}}]{VQITE}%
  \BibitemOpen
  \bibfield  {author} {\bibinfo {author} {\bibfnamefont {S.}~\bibnamefont {McArdle}}, \bibinfo {author} {\bibfnamefont {T.}~\bibnamefont {Jones}}, \bibinfo {author} {\bibfnamefont {S.}~\bibnamefont {Endo}}, \bibinfo {author} {\bibfnamefont {Y.}~\bibnamefont {Li}}, \bibinfo {author} {\bibfnamefont {S.~C.}\ \bibnamefont {Benjamin}},\ and\ \bibinfo {author} {\bibfnamefont {X.}~\bibnamefont {Yuan}},\ }\bibfield  {title} {\bibinfo {title} {Variational ansatz-based quantum simulation of imaginary time evolution},\ }\href {https://doi.org/10.1038/s41534-019-0187-2} {\bibfield  {journal} {\bibinfo  {journal} {npj Quantum Inf.}\ }\textbf {\bibinfo {volume} {5}},\ \bibinfo {pages} {75} (\bibinfo {year} {2019})}\BibitemShut {NoStop}%
\bibitem [{\citenamefont {Motta}\ \emph {et~al.}(2020)\citenamefont {Motta}, \citenamefont {Sun}, \citenamefont {Tan}, \citenamefont {O’Rourke}, \citenamefont {Ye}, \citenamefont {Minnich}, \citenamefont {Brand{\~a}o},\ and\ \citenamefont {Chan}}]{qite_chan20}%
  \BibitemOpen
  \bibfield  {author} {\bibinfo {author} {\bibfnamefont {M.}~\bibnamefont {Motta}}, \bibinfo {author} {\bibfnamefont {C.}~\bibnamefont {Sun}}, \bibinfo {author} {\bibfnamefont {A.~T.}\ \bibnamefont {Tan}}, \bibinfo {author} {\bibfnamefont {M.~J.}\ \bibnamefont {O’Rourke}}, \bibinfo {author} {\bibfnamefont {E.}~\bibnamefont {Ye}}, \bibinfo {author} {\bibfnamefont {A.~J.}\ \bibnamefont {Minnich}}, \bibinfo {author} {\bibfnamefont {F.~G.}\ \bibnamefont {Brand{\~a}o}},\ and\ \bibinfo {author} {\bibfnamefont {G.~K.-L.}\ \bibnamefont {Chan}},\ }\bibfield  {title} {\bibinfo {title} {Determining eigenstates and thermal states on a quantum computer using quantum imaginary time evolution},\ }\href {https://doi.org/10.1038/s41567-019-0704-4} {\bibfield  {journal} {\bibinfo  {journal} {Nat. Phys.}\ }\textbf {\bibinfo {volume} {16}},\ \bibinfo {pages} {205} (\bibinfo {year} {2020})}\BibitemShut {NoStop}%
\bibitem [{\citenamefont {Stokes}\ \emph {et~al.}(2020)\citenamefont {Stokes}, \citenamefont {Izaac}, \citenamefont {Killoran},\ and\ \citenamefont {Carleo}}]{stokes2020quantum}%
  \BibitemOpen
  \bibfield  {author} {\bibinfo {author} {\bibfnamefont {J.}~\bibnamefont {Stokes}}, \bibinfo {author} {\bibfnamefont {J.}~\bibnamefont {Izaac}}, \bibinfo {author} {\bibfnamefont {N.}~\bibnamefont {Killoran}},\ and\ \bibinfo {author} {\bibfnamefont {G.}~\bibnamefont {Carleo}},\ }\bibfield  {title} {\bibinfo {title} {Quantum natural gradient},\ }\href {https://doi.org/10.22331/q-2020-05-25-269} {\bibfield  {journal} {\bibinfo  {journal} {Quantum}\ }\textbf {\bibinfo {volume} {4}},\ \bibinfo {pages} {269} (\bibinfo {year} {2020})}\BibitemShut {NoStop}%
\bibitem [{\citenamefont {Grimsley}\ \emph {et~al.}(2019)\citenamefont {Grimsley}, \citenamefont {Economou}, \citenamefont {Barnes},\ and\ \citenamefont {Mayhall}}]{grimsleyAdaptiveVariationalAlgorithm2019}%
  \BibitemOpen
  \bibfield  {author} {\bibinfo {author} {\bibfnamefont {H.~R.}\ \bibnamefont {Grimsley}}, \bibinfo {author} {\bibfnamefont {S.~E.}\ \bibnamefont {Economou}}, \bibinfo {author} {\bibfnamefont {E.}~\bibnamefont {Barnes}},\ and\ \bibinfo {author} {\bibfnamefont {N.~J.}\ \bibnamefont {Mayhall}},\ }\bibfield  {title} {\bibinfo {title} {An adaptive variational algorithm for exact molecular simulations on a quantum computer},\ }\href {https://doi.org/10.1038/s41467-019-10988-2} {\bibfield  {journal} {\bibinfo  {journal} {Nat. Commun.}\ }\textbf {\bibinfo {volume} {10}},\ \bibinfo {pages} {3007} (\bibinfo {year} {2019})}\BibitemShut {NoStop}%
\bibitem [{\citenamefont {Tang}\ \emph {et~al.}(2021)\citenamefont {Tang}, \citenamefont {Shkolnikov}, \citenamefont {Barron}, \citenamefont {Grimsley}, \citenamefont {Mayhall}, \citenamefont {Barnes},\ and\ \citenamefont {Economou}}]{MayhallQubitAVQE}%
  \BibitemOpen
  \bibfield  {author} {\bibinfo {author} {\bibfnamefont {H.~L.}\ \bibnamefont {Tang}}, \bibinfo {author} {\bibfnamefont {V.}~\bibnamefont {Shkolnikov}}, \bibinfo {author} {\bibfnamefont {G.~S.}\ \bibnamefont {Barron}}, \bibinfo {author} {\bibfnamefont {H.~R.}\ \bibnamefont {Grimsley}}, \bibinfo {author} {\bibfnamefont {N.~J.}\ \bibnamefont {Mayhall}}, \bibinfo {author} {\bibfnamefont {E.}~\bibnamefont {Barnes}},\ and\ \bibinfo {author} {\bibfnamefont {S.~E.}\ \bibnamefont {Economou}},\ }\bibfield  {title} {\bibinfo {title} {Qubit-adapt-vqe: An adaptive algorithm for constructing hardware-efficient ans\"atze on a quantum processor},\ }\href {https://doi.org/10.1103/PRXQuantum.2.020310} {\bibfield  {journal} {\bibinfo  {journal} {PRX Quantum}\ }\textbf {\bibinfo {volume} {2}},\ \bibinfo {pages} {020310} (\bibinfo {year} {2021})}\BibitemShut {NoStop}%
\bibitem [{\citenamefont {Gomes}\ \emph {et~al.}(2020)\citenamefont {Gomes}, \citenamefont {Zhang}, \citenamefont {Berthusen}, \citenamefont {Wang}, \citenamefont {Ho}, \citenamefont {Orth},\ and\ \citenamefont {Yao}}]{smqite}%
  \BibitemOpen
  \bibfield  {author} {\bibinfo {author} {\bibfnamefont {N.}~\bibnamefont {Gomes}}, \bibinfo {author} {\bibfnamefont {F.}~\bibnamefont {Zhang}}, \bibinfo {author} {\bibfnamefont {N.~F.}\ \bibnamefont {Berthusen}}, \bibinfo {author} {\bibfnamefont {C.-Z.}\ \bibnamefont {Wang}}, \bibinfo {author} {\bibfnamefont {K.-M.}\ \bibnamefont {Ho}}, \bibinfo {author} {\bibfnamefont {P.~P.}\ \bibnamefont {Orth}},\ and\ \bibinfo {author} {\bibfnamefont {Y.-X.}\ \bibnamefont {Yao}},\ }\bibfield  {title} {\bibinfo {title} {Efficient step-merged quantum imaginary time evolution algorithm for quantum chemistry},\ }\href {https://doi.org/10.1021/acs.jctc.0c00666} {\bibfield  {journal} {\bibinfo  {journal} {J. Chem. Theory Comput.}\ }\textbf {\bibinfo {volume} {16}},\ \bibinfo {pages} {6256} (\bibinfo {year} {2020})}\BibitemShut {NoStop}%
\bibitem [{\citenamefont {Alipanah}\ \emph {et~al.}(2025)\citenamefont {Alipanah}, \citenamefont {Zhang}, \citenamefont {Yao}, \citenamefont {Thompson}, \citenamefont {Nguyen}, \citenamefont {Liu}, \citenamefont {Givi}, \citenamefont {McDermott},\ and\ \citenamefont {Mendoza-Arenas}}]{Alipanah2025QuantumQS}%
  \BibitemOpen
  \bibfield  {author} {\bibinfo {author} {\bibfnamefont {H.}~\bibnamefont {Alipanah}}, \bibinfo {author} {\bibfnamefont {F.}~\bibnamefont {Zhang}}, \bibinfo {author} {\bibfnamefont {Y.}~\bibnamefont {Yao}}, \bibinfo {author} {\bibfnamefont {R.}~\bibnamefont {Thompson}}, \bibinfo {author} {\bibfnamefont {N.}~\bibnamefont {Nguyen}}, \bibinfo {author} {\bibfnamefont {J.}~\bibnamefont {Liu}}, \bibinfo {author} {\bibfnamefont {P.}~\bibnamefont {Givi}}, \bibinfo {author} {\bibfnamefont {B.~J.}\ \bibnamefont {McDermott}},\ and\ \bibinfo {author} {\bibfnamefont {J.~J.}\ \bibnamefont {Mendoza-Arenas}},\ }\href {https://doi.org/10.48550/arXiv.2503.13729} {\bibinfo {title} {Quantum dynamics simulation of the advection-diffusion equation}} (\bibinfo {year} {2025}),\ \Eprint {https://arxiv.org/abs/2503.13729} {arXiv:2503.13729 [quant-ph]} \BibitemShut {NoStop}%
\bibitem [{\citenamefont {Zhang}\ \emph {et~al.}(2025)\citenamefont {Zhang}, \citenamefont {Wang}, \citenamefont {Iadecola}, \citenamefont {Orth},\ and\ \citenamefont {Yao}}]{Zhang2025AdaptiveVQ}%
  \BibitemOpen
  \bibfield  {author} {\bibinfo {author} {\bibfnamefont {F.}~\bibnamefont {Zhang}}, \bibinfo {author} {\bibfnamefont {C.-Z.}\ \bibnamefont {Wang}}, \bibinfo {author} {\bibfnamefont {T.}~\bibnamefont {Iadecola}}, \bibinfo {author} {\bibfnamefont {P.~P.}\ \bibnamefont {Orth}},\ and\ \bibinfo {author} {\bibfnamefont {Y.-X.}\ \bibnamefont {Yao}},\ }\bibfield  {title} {\bibinfo {title} {Adaptive variational quantum dynamics simulations with compressed circuits and fewer measurements},\ }\href {https://doi.org/10.1103/PhysRevB.111.094310} {\bibfield  {journal} {\bibinfo  {journal} {Phys. Rev. B}\ }\textbf {\bibinfo {volume} {111}},\ \bibinfo {pages} {094310} (\bibinfo {year} {2025})}\BibitemShut {NoStop}%
\bibitem [{\citenamefont {Anastasiou}\ \emph {et~al.}(2024)\citenamefont {Anastasiou}, \citenamefont {Chen}, \citenamefont {Mayhall}, \citenamefont {Barnes},\ and\ \citenamefont {Economou}}]{Anastasiou2024TETRIS}%
  \BibitemOpen
  \bibfield  {author} {\bibinfo {author} {\bibfnamefont {P.~G.}\ \bibnamefont {Anastasiou}}, \bibinfo {author} {\bibfnamefont {Y.}~\bibnamefont {Chen}}, \bibinfo {author} {\bibfnamefont {N.~J.}\ \bibnamefont {Mayhall}}, \bibinfo {author} {\bibfnamefont {E.}~\bibnamefont {Barnes}},\ and\ \bibinfo {author} {\bibfnamefont {S.~E.}\ \bibnamefont {Economou}},\ }\bibfield  {title} {\bibinfo {title} {Tetris-adapt-vqe: An adaptive algorithm that yields shallower, denser circuit ans\"atze},\ }\href {https://doi.org/10.1103/PhysRevResearch.6.013254} {\bibfield  {journal} {\bibinfo  {journal} {Phys. Rev. Res.}\ }\textbf {\bibinfo {volume} {6}},\ \bibinfo {pages} {013254} (\bibinfo {year} {2024})}\BibitemShut {NoStop}%
\bibitem [{\citenamefont {\ifmmode~\check{Z}\else \v{Z}\fi{}itko}\ and\ \citenamefont {Pruschke}(2009)}]{Zitko2009NRG}%
  \BibitemOpen
  \bibfield  {author} {\bibinfo {author} {\bibfnamefont {R.}~\bibnamefont {\ifmmode~\check{Z}\else \v{Z}\fi{}itko}}\ and\ \bibinfo {author} {\bibfnamefont {T.}~\bibnamefont {Pruschke}},\ }\bibfield  {title} {\bibinfo {title} {Energy resolution and discretization artifacts in the numerical renormalization group},\ }\href {https://doi.org/10.1103/PhysRevB.79.085106} {\bibfield  {journal} {\bibinfo  {journal} {Phys. Rev. B}\ }\textbf {\bibinfo {volume} {79}},\ \bibinfo {pages} {085106} (\bibinfo {year} {2009})}\BibitemShut {NoStop}%
\bibitem [{\citenamefont {Zitko}(2021)}]{zitko_2021_NRG}%
  \BibitemOpen
  \bibfield  {author} {\bibinfo {author} {\bibfnamefont {R.}~\bibnamefont {Zitko}},\ }\href {https://doi.org/10.5281/zenodo.4841076} {\bibinfo {title} {Nrg ljubljana}} (\bibinfo {year} {2021})\BibitemShut {NoStop}%
\bibitem [{\citenamefont {Parcollet}\ \emph {et~al.}(2015)\citenamefont {Parcollet}, \citenamefont {Ferrero}, \citenamefont {Ayral}, \citenamefont {Hafermann}, \citenamefont {Krivenko}, \citenamefont {Messio},\ and\ \citenamefont {Seth}}]{TRIQS2015}%
  \BibitemOpen
  \bibfield  {author} {\bibinfo {author} {\bibfnamefont {O.}~\bibnamefont {Parcollet}}, \bibinfo {author} {\bibfnamefont {M.}~\bibnamefont {Ferrero}}, \bibinfo {author} {\bibfnamefont {T.}~\bibnamefont {Ayral}}, \bibinfo {author} {\bibfnamefont {H.}~\bibnamefont {Hafermann}}, \bibinfo {author} {\bibfnamefont {I.}~\bibnamefont {Krivenko}}, \bibinfo {author} {\bibfnamefont {L.}~\bibnamefont {Messio}},\ and\ \bibinfo {author} {\bibfnamefont {P.}~\bibnamefont {Seth}},\ }\bibfield  {title} {\bibinfo {title} {{TRIQS: A toolbox for research on interacting quantum systems}},\ }\href {https://doi.org/10.1016/j.cpc.2015.04.023} {\bibfield  {journal} {\bibinfo  {journal} {Computer Physics Communications}\ }\textbf {\bibinfo {volume} {196}},\ \bibinfo {pages} {398} (\bibinfo {year} {2015})},\ \Eprint {https://arxiv.org/abs/1504.01952} {1504.01952} \BibitemShut {NoStop}%
\bibitem [{\citenamefont {Khindanov}\ \emph {et~al.}(2025)\citenamefont {Khindanov}, \citenamefont {Yao},\ and\ \citenamefont {Iadecola}}]{Khindanov2025}%
  \BibitemOpen
  \bibfield  {author} {\bibinfo {author} {\bibfnamefont {A.}~\bibnamefont {Khindanov}}, \bibinfo {author} {\bibfnamefont {Y.}~\bibnamefont {Yao}},\ and\ \bibinfo {author} {\bibfnamefont {T.}~\bibnamefont {Iadecola}},\ }\bibfield  {title} {\bibinfo {title} {Robust preparation of ground state phases under noisy imaginary time evolution},\ }\href {https://doi.org/10.1103/PhysRevResearch.7.013263} {\bibfield  {journal} {\bibinfo  {journal} {Phys. Rev. Res.}\ }\textbf {\bibinfo {volume} {7}},\ \bibinfo {pages} {013263} (\bibinfo {year} {2025})}\BibitemShut {NoStop}%
\bibitem [{\citenamefont {Getelina}\ \emph {et~al.}(2023)\citenamefont {Getelina}, \citenamefont {Gomes}, \citenamefont {Iadecola}, \citenamefont {Orth},\ and\ \citenamefont {Yao}}]{getelina2023adaptive}%
  \BibitemOpen
  \bibfield  {author} {\bibinfo {author} {\bibfnamefont {J.~C.}\ \bibnamefont {Getelina}}, \bibinfo {author} {\bibfnamefont {N.}~\bibnamefont {Gomes}}, \bibinfo {author} {\bibfnamefont {T.}~\bibnamefont {Iadecola}}, \bibinfo {author} {\bibfnamefont {P.~P.}\ \bibnamefont {Orth}},\ and\ \bibinfo {author} {\bibfnamefont {Y.-X.}\ \bibnamefont {Yao}},\ }\bibfield  {title} {\bibinfo {title} {{Adaptive variational quantum minimally entangled typical thermal states for finite temperature simulations}},\ }\href {https://doi.org/10.21468/SciPostPhys.15.3.102} {\bibfield  {journal} {\bibinfo  {journal} {SciPost Phys.}\ }\textbf {\bibinfo {volume} {15}},\ \bibinfo {pages} {102} (\bibinfo {year} {2023})}\BibitemShut {NoStop}%
\bibitem [{\citenamefont {Self}\ \emph {et~al.}(2024)\citenamefont {Self}, \citenamefont {Benedetti},\ and\ \citenamefont {Amaro}}]{self2024protecting}%
  \BibitemOpen
  \bibfield  {author} {\bibinfo {author} {\bibfnamefont {C.~N.}\ \bibnamefont {Self}}, \bibinfo {author} {\bibfnamefont {M.}~\bibnamefont {Benedetti}},\ and\ \bibinfo {author} {\bibfnamefont {D.}~\bibnamefont {Amaro}},\ }\bibfield  {title} {\bibinfo {title} {Protecting expressive circuits with a quantum error detection code},\ }\href {https://doi.org/10.1038/s41567-023-02282-2} {\bibfield  {journal} {\bibinfo  {journal} {Nature Physics}\ }\textbf {\bibinfo {volume} {20}},\ \bibinfo {pages} {219} (\bibinfo {year} {2024})}\BibitemShut {NoStop}%
\bibitem [{\citenamefont {Krinner}\ \emph {et~al.}(2022)\citenamefont {Krinner}, \citenamefont {Lacroix}, \citenamefont {Remm}, \citenamefont {Di~Paolo}, \citenamefont {Genois}, \citenamefont {Leroux}, \citenamefont {Hellings}, \citenamefont {Lazar}, \citenamefont {Swiadek}, \citenamefont {Herrmann} \emph {et~al.}}]{krinner2022realizing}%
  \BibitemOpen
  \bibfield  {author} {\bibinfo {author} {\bibfnamefont {S.}~\bibnamefont {Krinner}}, \bibinfo {author} {\bibfnamefont {N.}~\bibnamefont {Lacroix}}, \bibinfo {author} {\bibfnamefont {A.}~\bibnamefont {Remm}}, \bibinfo {author} {\bibfnamefont {A.}~\bibnamefont {Di~Paolo}}, \bibinfo {author} {\bibfnamefont {E.}~\bibnamefont {Genois}}, \bibinfo {author} {\bibfnamefont {C.}~\bibnamefont {Leroux}}, \bibinfo {author} {\bibfnamefont {C.}~\bibnamefont {Hellings}}, \bibinfo {author} {\bibfnamefont {S.}~\bibnamefont {Lazar}}, \bibinfo {author} {\bibfnamefont {F.}~\bibnamefont {Swiadek}}, \bibinfo {author} {\bibfnamefont {J.}~\bibnamefont {Herrmann}}, \emph {et~al.},\ }\bibfield  {title} {\bibinfo {title} {Realizing repeated quantum error correction in a distance-three surface code},\ }\href {https://doi.org/10.1038/s41586-022-04566-8} {\bibfield  {journal} {\bibinfo  {journal} {Nature}\ }\textbf {\bibinfo {volume} {605}},\ \bibinfo {pages} {669} (\bibinfo {year} {2022})}\BibitemShut {NoStop}%
\bibitem [{\citenamefont {Sivak}\ \emph {et~al.}(2023)\citenamefont {Sivak}, \citenamefont {Eickbusch}, \citenamefont {Royer}, \citenamefont {Singh}, \citenamefont {Tsioutsios}, \citenamefont {Ganjam}, \citenamefont {Miano}, \citenamefont {Brock}, \citenamefont {Ding}, \citenamefont {Frunzio} \emph {et~al.}}]{sivak2023real}%
  \BibitemOpen
  \bibfield  {author} {\bibinfo {author} {\bibfnamefont {V.~V.}\ \bibnamefont {Sivak}}, \bibinfo {author} {\bibfnamefont {A.}~\bibnamefont {Eickbusch}}, \bibinfo {author} {\bibfnamefont {B.}~\bibnamefont {Royer}}, \bibinfo {author} {\bibfnamefont {S.}~\bibnamefont {Singh}}, \bibinfo {author} {\bibfnamefont {I.}~\bibnamefont {Tsioutsios}}, \bibinfo {author} {\bibfnamefont {S.}~\bibnamefont {Ganjam}}, \bibinfo {author} {\bibfnamefont {A.}~\bibnamefont {Miano}}, \bibinfo {author} {\bibfnamefont {B.}~\bibnamefont {Brock}}, \bibinfo {author} {\bibfnamefont {A.}~\bibnamefont {Ding}}, \bibinfo {author} {\bibfnamefont {L.}~\bibnamefont {Frunzio}}, \emph {et~al.},\ }\bibfield  {title} {\bibinfo {title} {Real-time quantum error correction beyond break-even},\ }\href {https://doi.org/10.1038/s41586-023-05782-6} {\bibfield  {journal} {\bibinfo  {journal} {Nature}\ }\textbf {\bibinfo {volume} {616}},\ \bibinfo {pages} {50} (\bibinfo {year} {2023})}\BibitemShut {NoStop}%
\bibitem [{\citenamefont {Bluvstein}\ \emph {et~al.}(2024)\citenamefont {Bluvstein}, \citenamefont {Evered}, \citenamefont {Geim}, \citenamefont {Li}, \citenamefont {Zhou}, \citenamefont {Manovitz}, \citenamefont {Ebadi}, \citenamefont {Cain}, \citenamefont {Kalinowski}, \citenamefont {Hangleiter} \emph {et~al.}}]{bluvstein2024logical}%
  \BibitemOpen
  \bibfield  {author} {\bibinfo {author} {\bibfnamefont {D.}~\bibnamefont {Bluvstein}}, \bibinfo {author} {\bibfnamefont {S.~J.}\ \bibnamefont {Evered}}, \bibinfo {author} {\bibfnamefont {A.~A.}\ \bibnamefont {Geim}}, \bibinfo {author} {\bibfnamefont {S.~H.}\ \bibnamefont {Li}}, \bibinfo {author} {\bibfnamefont {H.}~\bibnamefont {Zhou}}, \bibinfo {author} {\bibfnamefont {T.}~\bibnamefont {Manovitz}}, \bibinfo {author} {\bibfnamefont {S.}~\bibnamefont {Ebadi}}, \bibinfo {author} {\bibfnamefont {M.}~\bibnamefont {Cain}}, \bibinfo {author} {\bibfnamefont {M.}~\bibnamefont {Kalinowski}}, \bibinfo {author} {\bibfnamefont {D.}~\bibnamefont {Hangleiter}}, \emph {et~al.},\ }\bibfield  {title} {\bibinfo {title} {Logical quantum processor based on reconfigurable atom arrays},\ }\href {https://doi.org/10.1038/s41586-023-06927-3} {\bibfield  {journal} {\bibinfo  {journal} {Nature}\ }\textbf {\bibinfo {volume} {626}},\ \bibinfo {pages} {58} (\bibinfo {year} {2024})}\BibitemShut {NoStop}%
\bibitem [{\citenamefont {Paetznick}\ \emph {et~al.}(2024)\citenamefont {Paetznick}, \citenamefont {da~Silva}, \citenamefont {Ryan-Anderson}, \citenamefont {Bello-Rivas}, \citenamefont {Campora~III}, \citenamefont {Chernoguzov}, \citenamefont {Dreiling}, \citenamefont {Foltz}, \citenamefont {Frachon}, \citenamefont {Gaebler} \emph {et~al.}}]{paetznick2024demonstration}%
  \BibitemOpen
  \bibfield  {author} {\bibinfo {author} {\bibfnamefont {A.}~\bibnamefont {Paetznick}}, \bibinfo {author} {\bibfnamefont {M.}~\bibnamefont {da~Silva}}, \bibinfo {author} {\bibfnamefont {C.}~\bibnamefont {Ryan-Anderson}}, \bibinfo {author} {\bibfnamefont {J.}~\bibnamefont {Bello-Rivas}}, \bibinfo {author} {\bibfnamefont {J.}~\bibnamefont {Campora~III}}, \bibinfo {author} {\bibfnamefont {A.}~\bibnamefont {Chernoguzov}}, \bibinfo {author} {\bibfnamefont {J.}~\bibnamefont {Dreiling}}, \bibinfo {author} {\bibfnamefont {C.}~\bibnamefont {Foltz}}, \bibinfo {author} {\bibfnamefont {F.}~\bibnamefont {Frachon}}, \bibinfo {author} {\bibfnamefont {J.}~\bibnamefont {Gaebler}}, \emph {et~al.},\ }\bibfield  {title} {\bibinfo {title} {Demonstration of logical qubits and repeated error correction with better-than-physical error rates},\ }\bibfield  {journal} {\bibinfo  {journal} {arXiv preprint arXiv:2404.02280}\ }\href {https://doi.org/10.48550/arXiv.2404.02280} {10.48550/arXiv.2404.02280} (\bibinfo {year}
  {2024})\BibitemShut {NoStop}%
\bibitem [{\citenamefont {Gupta}\ \emph {et~al.}(2024)\citenamefont {Gupta}, \citenamefont {Sundaresan}, \citenamefont {Alexander}, \citenamefont {Wood}, \citenamefont {Merkel}, \citenamefont {Healy}, \citenamefont {Hillenbrand}, \citenamefont {Jochym-O’Connor}, \citenamefont {Wootton}, \citenamefont {Yoder} \emph {et~al.}}]{gupta2024encoding}%
  \BibitemOpen
  \bibfield  {author} {\bibinfo {author} {\bibfnamefont {R.~S.}\ \bibnamefont {Gupta}}, \bibinfo {author} {\bibfnamefont {N.}~\bibnamefont {Sundaresan}}, \bibinfo {author} {\bibfnamefont {T.}~\bibnamefont {Alexander}}, \bibinfo {author} {\bibfnamefont {C.~J.}\ \bibnamefont {Wood}}, \bibinfo {author} {\bibfnamefont {S.~T.}\ \bibnamefont {Merkel}}, \bibinfo {author} {\bibfnamefont {M.~B.}\ \bibnamefont {Healy}}, \bibinfo {author} {\bibfnamefont {M.}~\bibnamefont {Hillenbrand}}, \bibinfo {author} {\bibfnamefont {T.}~\bibnamefont {Jochym-O’Connor}}, \bibinfo {author} {\bibfnamefont {J.~R.}\ \bibnamefont {Wootton}}, \bibinfo {author} {\bibfnamefont {T.~J.}\ \bibnamefont {Yoder}}, \emph {et~al.},\ }\bibfield  {title} {\bibinfo {title} {Encoding a magic state with beyond break-even fidelity},\ }\href {https://doi.org/10.1038/s41586-023-06846-3} {\bibfield  {journal} {\bibinfo  {journal} {Nature}\ }\textbf {\bibinfo {volume} {625}},\ \bibinfo {pages} {259} (\bibinfo {year} {2024})}\BibitemShut {NoStop}%
\bibitem [{\citenamefont {AI}\ \emph {et~al.}(2024)\citenamefont {AI} \emph {et~al.}}]{AI2024QECBelow}%
  \BibitemOpen
  \bibfield  {author} {\bibinfo {author} {\bibfnamefont {G.~Q.}\ \bibnamefont {AI}} \emph {et~al.},\ }\bibfield  {title} {\bibinfo {title} {Quantum error correction below the surface code threshold},\ }\href {https://doi.org/10.1038/s41586-024-08449-y} {\bibfield  {journal} {\bibinfo  {journal} {Nature}\ }\textbf {\bibinfo {volume} {638}},\ \bibinfo {pages} {920} (\bibinfo {year} {2024})}\BibitemShut {NoStop}%
\bibitem [{\citenamefont {Knill}(2004{\natexlab{a}})}]{Knill2004schemes}%
  \BibitemOpen
  \bibfield  {author} {\bibinfo {author} {\bibfnamefont {E.}~\bibnamefont {Knill}},\ }\href {https://arxiv.org/abs/quant-ph/0402171} {\bibinfo {title} {Fault-tolerant postselected quantum computation: Schemes}} (\bibinfo {year} {2004}{\natexlab{a}}),\ \Eprint {https://arxiv.org/abs/quant-ph/0402171} {arXiv:quant-ph/0402171 [quant-ph]} \BibitemShut {NoStop}%
\bibitem [{\citenamefont {Knill}(2004{\natexlab{b}})}]{Knill2004threshold}%
  \BibitemOpen
  \bibfield  {author} {\bibinfo {author} {\bibfnamefont {E.}~\bibnamefont {Knill}},\ }\href {https://arxiv.org/abs/quant-ph/0404104} {\bibinfo {title} {Fault-tolerant postselected quantum computation: Threshold analysis}} (\bibinfo {year} {2004}{\natexlab{b}}),\ \Eprint {https://arxiv.org/abs/quant-ph/0404104} {arXiv:quant-ph/0404104 [quant-ph]} \BibitemShut {NoStop}%
\bibitem [{\citenamefont {Gowrishankar}\ \emph {et~al.}(2024)\citenamefont {Gowrishankar}, \citenamefont {Claudino}, \citenamefont {Wright},\ and\ \citenamefont {Humble}}]{gowrishankar2024logical}%
  \BibitemOpen
  \bibfield  {author} {\bibinfo {author} {\bibfnamefont {M.}~\bibnamefont {Gowrishankar}}, \bibinfo {author} {\bibfnamefont {D.}~\bibnamefont {Claudino}}, \bibinfo {author} {\bibfnamefont {J.}~\bibnamefont {Wright}},\ and\ \bibinfo {author} {\bibfnamefont {T.}~\bibnamefont {Humble}},\ }\bibfield  {title} {\bibinfo {title} {Logical error rates for a [[4, 2, 2]]-encoded variational quantum eigensolver ansatz},\ }\bibfield  {journal} {\bibinfo  {journal} {arXiv preprint arXiv:2405.03032}\ }\href {https://doi.org/10.48550/arXiv.2405.03032} {10.48550/arXiv.2405.03032} (\bibinfo {year} {2024})\BibitemShut {NoStop}%
\bibitem [{\citenamefont {He}\ \emph {et~al.}(2024)\citenamefont {He}, \citenamefont {Amaro}, \citenamefont {Shaydulin},\ and\ \citenamefont {Pistoia}}]{he2024performance}%
  \BibitemOpen
  \bibfield  {author} {\bibinfo {author} {\bibfnamefont {Z.}~\bibnamefont {He}}, \bibinfo {author} {\bibfnamefont {D.}~\bibnamefont {Amaro}}, \bibinfo {author} {\bibfnamefont {R.}~\bibnamefont {Shaydulin}},\ and\ \bibinfo {author} {\bibfnamefont {M.}~\bibnamefont {Pistoia}},\ }\bibfield  {title} {\bibinfo {title} {Performance of quantum approximate optimization with quantum error detection},\ }\bibfield  {journal} {\bibinfo  {journal} {arXiv preprint arXiv:2409.12104}\ }\href {https://doi.org/10.48550/arXiv.2409.12104} {10.48550/arXiv.2409.12104} (\bibinfo {year} {2024})\BibitemShut {NoStop}%
\bibitem [{\citenamefont {Chao}\ and\ \citenamefont {Reichardt}(2018)}]{Chao2018FaultTQC}%
  \BibitemOpen
  \bibfield  {author} {\bibinfo {author} {\bibfnamefont {R.}~\bibnamefont {Chao}}\ and\ \bibinfo {author} {\bibfnamefont {B.~W.}\ \bibnamefont {Reichardt}},\ }\bibfield  {title} {\bibinfo {title} {Fault-tolerant quantum computation with few qubits},\ }\bibfield  {journal} {\bibinfo  {journal} {npj Quantum Information}\ }\textbf {\bibinfo {volume} {4}},\ \href {https://doi.org/10.1038/s41534-018-0085-z} {10.1038/s41534-018-0085-z} (\bibinfo {year} {2018})\BibitemShut {NoStop}%
\bibitem [{\citenamefont {Abraham}\ \emph {et~al.}(2019)\citenamefont {Abraham}, \citenamefont {Akhalwaya}, \citenamefont {Aleksandrowicz}, \citenamefont {Alexander}, \citenamefont {Alexandrowics}, \citenamefont {Arbel}, \citenamefont {Asfaw}, \citenamefont {Azaustre}, \citenamefont {AzizNgoueya}, \citenamefont {Barkoutsos}, \citenamefont {Barron}, \citenamefont {Bello}, \citenamefont {Ben-Haim}, \citenamefont {Bevenius} \emph {et~al.}}]{Qiskit}%
  \BibitemOpen
  \bibfield  {author} {\bibinfo {author} {\bibfnamefont {H.}~\bibnamefont {Abraham}}, \bibinfo {author} {\bibfnamefont {I.~Y.}\ \bibnamefont {Akhalwaya}}, \bibinfo {author} {\bibfnamefont {G.}~\bibnamefont {Aleksandrowicz}}, \bibinfo {author} {\bibfnamefont {T.}~\bibnamefont {Alexander}}, \bibinfo {author} {\bibfnamefont {G.}~\bibnamefont {Alexandrowics}}, \bibinfo {author} {\bibfnamefont {E.}~\bibnamefont {Arbel}}, \bibinfo {author} {\bibfnamefont {A.}~\bibnamefont {Asfaw}}, \bibinfo {author} {\bibfnamefont {C.}~\bibnamefont {Azaustre}}, \bibinfo {author} {\bibnamefont {AzizNgoueya}}, \bibinfo {author} {\bibfnamefont {P.}~\bibnamefont {Barkoutsos}}, \bibinfo {author} {\bibfnamefont {G.}~\bibnamefont {Barron}}, \bibinfo {author} {\bibfnamefont {L.}~\bibnamefont {Bello}}, \bibinfo {author} {\bibfnamefont {Y.}~\bibnamefont {Ben-Haim}}, \bibinfo {author} {\bibfnamefont {D.}~\bibnamefont {Bevenius}}, \emph {et~al.},\ }\href {https://doi.org/10.5281/zenodo.2562111} {\bibinfo {title} {Qiskit: An open-source
  framework for quantum computing}} (\bibinfo {year} {2019})\BibitemShut {NoStop}%
\bibitem [{\citenamefont {{Getelina}}\ \emph {et~al.}(2024)\citenamefont {{Getelina}}, \citenamefont {{Sharma}}, \citenamefont {{Iadecola}}, \citenamefont {{Orth}},\ and\ \citenamefont {{Yao}}}]{Getelina2024QuantumSE}%
  \BibitemOpen
  \bibfield  {author} {\bibinfo {author} {\bibfnamefont {J.~C.}\ \bibnamefont {{Getelina}}}, \bibinfo {author} {\bibfnamefont {P.}~\bibnamefont {{Sharma}}}, \bibinfo {author} {\bibfnamefont {T.}~\bibnamefont {{Iadecola}}}, \bibinfo {author} {\bibfnamefont {P.~P.}\ \bibnamefont {{Orth}}},\ and\ \bibinfo {author} {\bibfnamefont {Y.-X.}\ \bibnamefont {{Yao}}},\ }\bibfield  {title} {\bibinfo {title} {{Quantum subspace expansion in the presence of hardware noise}},\ }\href {https://doi.org/10.48550/arXiv.2404.09132} {\bibfield  {journal} {\bibinfo  {journal} {arXiv e-prints}\ ,\ \bibinfo {eid} {arXiv:2404.09132}} (\bibinfo {year} {2024})},\ \Eprint {https://arxiv.org/abs/2404.09132} {arXiv:2404.09132} \BibitemShut {NoStop}%
\bibitem [{\citenamefont {Kanno}\ \emph {et~al.}(2023)\citenamefont {Kanno}, \citenamefont {Kohda}, \citenamefont {Imai}, \citenamefont {Koh}, \citenamefont {Mitarai}, \citenamefont {Mizukami},\ and\ \citenamefont {Nakagawa}}]{kanno2023quantum}%
  \BibitemOpen
  \bibfield  {author} {\bibinfo {author} {\bibfnamefont {K.}~\bibnamefont {Kanno}}, \bibinfo {author} {\bibfnamefont {M.}~\bibnamefont {Kohda}}, \bibinfo {author} {\bibfnamefont {R.}~\bibnamefont {Imai}}, \bibinfo {author} {\bibfnamefont {S.}~\bibnamefont {Koh}}, \bibinfo {author} {\bibfnamefont {K.}~\bibnamefont {Mitarai}}, \bibinfo {author} {\bibfnamefont {W.}~\bibnamefont {Mizukami}},\ and\ \bibinfo {author} {\bibfnamefont {Y.~O.}\ \bibnamefont {Nakagawa}},\ }\href {https://doi.org/10.48550/arXiv.2302.11320} {\bibinfo {title} {Quantum-selected configuration interaction: classical diagonalization of hamiltonians in subspaces selected by quantum computers}} (\bibinfo {year} {2023}),\ \Eprint {https://arxiv.org/abs/2302.11320} {arXiv:2302.11320 [quant-ph]} \BibitemShut {NoStop}%
\bibitem [{\citenamefont {Robledo-Moreno}\ \emph {et~al.}(2024)\citenamefont {Robledo-Moreno}, \citenamefont {Motta}, \citenamefont {Haas}, \citenamefont {Javadi-Abhari}, \citenamefont {Jurcevic}, \citenamefont {Kirby}, \citenamefont {Martiel}, \citenamefont {Sharma}, \citenamefont {Sharma}, \citenamefont {Shirakawa} \emph {et~al.}}]{robledo2024chemistry}%
  \BibitemOpen
  \bibfield  {author} {\bibinfo {author} {\bibfnamefont {J.}~\bibnamefont {Robledo-Moreno}}, \bibinfo {author} {\bibfnamefont {M.}~\bibnamefont {Motta}}, \bibinfo {author} {\bibfnamefont {H.}~\bibnamefont {Haas}}, \bibinfo {author} {\bibfnamefont {A.}~\bibnamefont {Javadi-Abhari}}, \bibinfo {author} {\bibfnamefont {P.}~\bibnamefont {Jurcevic}}, \bibinfo {author} {\bibfnamefont {W.}~\bibnamefont {Kirby}}, \bibinfo {author} {\bibfnamefont {S.}~\bibnamefont {Martiel}}, \bibinfo {author} {\bibfnamefont {K.}~\bibnamefont {Sharma}}, \bibinfo {author} {\bibfnamefont {S.}~\bibnamefont {Sharma}}, \bibinfo {author} {\bibfnamefont {T.}~\bibnamefont {Shirakawa}}, \emph {et~al.},\ }\bibfield  {title} {\bibinfo {title} {Chemistry beyond exact solutions on a quantum-centric supercomputer},\ }\bibfield  {journal} {\bibinfo  {journal} {arXiv preprint arXiv:2405.05068}\ }\href {https://doi.org/10.48550/arXiv.2405.05068} {10.48550/arXiv.2405.05068} (\bibinfo {year} {2024})\BibitemShut {NoStop}%
\bibitem [{\citenamefont {Yu}\ \emph {et~al.}(2025)\citenamefont {Yu}, \citenamefont {Moreno}, \citenamefont {Iosue}, \citenamefont {Bertels}, \citenamefont {Claudino}, \citenamefont {Fuller}, \citenamefont {Groszkowski}, \citenamefont {Humble}, \citenamefont {Jurcevic}, \citenamefont {Kirby}, \citenamefont {Maier}, \citenamefont {Motta}, \citenamefont {Pokharel}, \citenamefont {Seif}, \citenamefont {Shehata}, \citenamefont {Sung}, \citenamefont {Tran}, \citenamefont {Tripathi}, \citenamefont {Mezzacapo},\ and\ \citenamefont {Sharma}}]{Yu2025QuantumCA}%
  \BibitemOpen
  \bibfield  {author} {\bibinfo {author} {\bibfnamefont {J.}~\bibnamefont {Yu}}, \bibinfo {author} {\bibfnamefont {J.~R.}\ \bibnamefont {Moreno}}, \bibinfo {author} {\bibfnamefont {J.~T.}\ \bibnamefont {Iosue}}, \bibinfo {author} {\bibfnamefont {L.}~\bibnamefont {Bertels}}, \bibinfo {author} {\bibfnamefont {D.}~\bibnamefont {Claudino}}, \bibinfo {author} {\bibfnamefont {B.}~\bibnamefont {Fuller}}, \bibinfo {author} {\bibfnamefont {P.}~\bibnamefont {Groszkowski}}, \bibinfo {author} {\bibfnamefont {T.~S.}\ \bibnamefont {Humble}}, \bibinfo {author} {\bibfnamefont {P.}~\bibnamefont {Jurcevic}}, \bibinfo {author} {\bibfnamefont {W.}~\bibnamefont {Kirby}}, \bibinfo {author} {\bibfnamefont {T.~A.}\ \bibnamefont {Maier}}, \bibinfo {author} {\bibfnamefont {M.}~\bibnamefont {Motta}}, \bibinfo {author} {\bibfnamefont {B.}~\bibnamefont {Pokharel}}, \bibinfo {author} {\bibfnamefont {A.}~\bibnamefont {Seif}}, \bibinfo {author} {\bibfnamefont {A.}~\bibnamefont {Shehata}}, \bibinfo {author} {\bibfnamefont {K.~J.}\
  \bibnamefont {Sung}}, \bibinfo {author} {\bibfnamefont {M.~C.}\ \bibnamefont {Tran}}, \bibinfo {author} {\bibfnamefont {V.}~\bibnamefont {Tripathi}}, \bibinfo {author} {\bibfnamefont {A.}~\bibnamefont {Mezzacapo}},\ and\ \bibinfo {author} {\bibfnamefont {K.}~\bibnamefont {Sharma}},\ }\href {https://doi.org/10.48550/arXiv.2501.09702} {\bibinfo {title} {Quantum-centric algorithm for sample-based krylov diagonalization}} (\bibinfo {year} {2025}),\ \Eprint {https://arxiv.org/abs/2501.09702} {arXiv:2501.09702 [quant-ph]} \BibitemShut {NoStop}%
\bibitem [{\citenamefont {Mikkelsen}\ and\ \citenamefont {Nakagawa}(2025)}]{Mikkelsen2025QuantumSCI}%
  \BibitemOpen
  \bibfield  {author} {\bibinfo {author} {\bibfnamefont {M.}~\bibnamefont {Mikkelsen}}\ and\ \bibinfo {author} {\bibfnamefont {Y.~O.}\ \bibnamefont {Nakagawa}},\ }\href {https://doi.org/10.48550/arXiv.2412.13839} {\bibinfo {title} {Quantum-selected configuration interaction with time-evolved state}} (\bibinfo {year} {2025}),\ \Eprint {https://arxiv.org/abs/2412.13839} {arXiv:2412.13839 [quant-ph]} \BibitemShut {NoStop}%
\bibitem [{\citenamefont {Schir{\'o}}\ and\ \citenamefont {Fabrizio}(2010)}]{schiro10tdga}%
  \BibitemOpen
  \bibfield  {author} {\bibinfo {author} {\bibfnamefont {M.}~\bibnamefont {Schir{\'o}}}\ and\ \bibinfo {author} {\bibfnamefont {M.}~\bibnamefont {Fabrizio}},\ }\bibfield  {title} {\bibinfo {title} {Time-dependent mean field theory for quench dynamics in correlated electron systems},\ }\href {https://doi.org/10.1103/PhysRevLett.105.076401} {\bibfield  {journal} {\bibinfo  {journal} {Phys. Rev. Lett.}\ }\textbf {\bibinfo {volume} {105}},\ \bibinfo {pages} {076401} (\bibinfo {year} {2010})}\BibitemShut {NoStop}%
\bibitem [{\citenamefont {Guerci}\ \emph {et~al.}(2023)\citenamefont {Guerci}, \citenamefont {Capone},\ and\ \citenamefont {Lanat\`a}}]{Guerci2023TDgGA}%
  \BibitemOpen
  \bibfield  {author} {\bibinfo {author} {\bibfnamefont {D.}~\bibnamefont {Guerci}}, \bibinfo {author} {\bibfnamefont {M.}~\bibnamefont {Capone}},\ and\ \bibinfo {author} {\bibfnamefont {N.}~\bibnamefont {Lanat\`a}},\ }\bibfield  {title} {\bibinfo {title} {Time-dependent ghost gutzwiller nonequilibrium dynamics},\ }\href {https://doi.org/10.1103/PhysRevResearch.5.L032023} {\bibfield  {journal} {\bibinfo  {journal} {Phys. Rev. Res.}\ }\textbf {\bibinfo {volume} {5}},\ \bibinfo {pages} {L032023} (\bibinfo {year} {2023})}\BibitemShut {NoStop}%
\end{thebibliography}%

\end{document}